\documentclass{new_tlp_no_date}
\usepackage{breakurl} 

\usepackage{hyperref}

\usepackage{url}

\usepackage{stmaryrd}
\usepackage{amsmath}
\usepackage{amssymb}
\usepackage{xspace}
\usepackage{booktabs}
\usepackage{subcaption}

\usepackage{graphicx}

\usepackage[pdftex,dvipsnames]{xcolor}
\usepackage{tikz}
\usetikzlibrary{positioning,shapes,arrows,fit,shadows,calc,matrix}

\usepackage{soul}
\definecolor{Light}{gray}{.90}
\sethlcolor{Light}
\newcommand{\hltexttt}[1]{\texttt{\small\hl{\mbox{#1}}}}
  
\usepackage{natbib}
\renewcommand{\cite}[1]{\citep{#1}}

\usepackage{placeins} 

\usepackage[inline]{enumitem}

\usepackage{listings}

\definecolor{ciaoframe}     {rgb}{  0,    0,  0.3}
\definecolor{ciaostring}    {rgb}{0.6, 0.46, 0.33}
\definecolor{ciaooperators} {rgb}{0.1, 0.15,  0.6}
\definecolor{ciaokeywords}  {rgb}{0.1, 0.15,  0.6}
\definecolor{ciaoassertions}{rgb}{0.1, 0.15,  0.6}
\definecolor{ciaotrust}     {RGB}{200, 130,     0}
\definecolor{ciaocheck}     {rgb}{0.1, 0.2,   0.8}
\definecolor{ciaochecked}   {rgb}{0.2, 0.34,  0.1}
\definecolor{ciaotrue}      {rgb}{0.2, 0.34,  0.1}
\definecolor{ciaofalse}     {rgb}{0.6,  0.0, 0.09}
\definecolor{ciaoprops}     {rgb}{0.1,  0.2,  0.8}
\definecolor{ciaocomment}   {rgb}{0.8,  0.3,  0.3}

\newcommand{\prettylstciao}[0]{
\lstset{language=Prolog,
  frameround=fttt,
  frame=ltrb,
  rulecolor=\color{ciaoframe},
  numbers=left,numberstyle=\tiny,stepnumber=1,numbersep=8pt,
  tabsize=4,
  breaklines=true,breakatwhitespace=true,
  basicstyle=\scriptsize\ttfamily, 
  showlines=true,
  showspaces=false,
  showtabs=false,
  escapechar=@,
  escapeinside={~~},
  commentstyle=\color{ciaocomment},
  stringstyle=\color{ciaostring},
  showstringspaces=false,
  deletekeywords={true}, 
  keywordstyle={\color{ciaooperators}\bfseries}, 
  classoffset=1, 
        otherkeywords={>,<,>=,=<,.,;,-,!,=,*,\&,+,:-,[,],|,->,:,:=,\#},
        keywordstyle={\color{ciaokeywords}\bfseries},
  classoffset=2,
       morekeywords={module,use_module,dynamic,export,import,multifile,impl_defined,trait,impl,mode},
       keywordstyle={\color{ciaokeywords}\bfseries},
       morekeywords={pred,prop,calls,success,comp},
       keywordstyle={\color{ciaoassertions}\bfseries},
  classoffset=4,
       morekeywords={trust,trust_default,entry},
       keywordstyle={\color{ciaotrust}\bfseries},
  classoffset=5,
       morekeywords={check},
       keywordstyle={\color{ciaocheck}\bfseries},
  classoffset=6,
       morekeywords={checked},
       keywordstyle={\color{ciaochecked}\bfseries},
  classoffset=7,
       morekeywords={true},
       keywordstyle={\color{ciaotrue}\bfseries},
  classoffset=8,
       morekeywords={false},
       keywordstyle={\color{ciaofalse}\bfseries},
  classoffset=9,
       morekeywords={even,nat,str,int,flt,atm,term,num,var,list,ground,mshare,
                    rsize,cardinality,not_fails,exp,cost,costb,steps_ub,steps_lb,
                    size_ub,size_lb,covered,mut_exclusive,head_cost,literal_cost,
                    is_det,non_det,length,terminates,steps_o,resource,socket,seff,string,
                    lowercase
       },
       keywordstyle={\color{ciaoprops}\bfseries},
  classoffset=0, 
}}

\newcommand{\pre}{\ensuremath{\mathit{Pre}}\xspace}
\newcommand{\post}{\ensuremath{\mathit{Post}}\xspace}
\newcommand{\comprops}{\ensuremath{\mathit{Comp}}\xspace}
\newcommand{\head}{\ensuremath{\mathit{Head}}\xspace}
\newcommand{\status}{\ensuremath{\mathit{Status}}\xspace}

\newcommand{\achecked}{\textcolor{ciaochecked}{checked}}
\newcommand{\acheck}{\textcolor{ciaocheck}{check}}
\newcommand{\afalse}{\textcolor{ciaofalse}{false}}
\newcommand{\atrue}{\textcolor{ciaotrue}{true}}

\newcommand{\pred}{\texttt{pred}\xspace}

\newcommand{\callsAsrN}{\textsf{calls}\xspace}
\newcommand{\callsAsr}[2]{\ensuremath{\callsAsrN(#1, #2)}}
\newcommand{\successAsrN}{\textsf{success}}
\newcommand{\successAsr}[3]{\ensuremath{\successAsrN(#1, #2, #3)}}
\newcommand{\compAsrN}{\textsf{comp}}

\newcommand{\compAsr}[3]{\ensuremath{\compAsrN(#1, #2, #3)}}

\def\tuple#1{\langle #1 \rangle}
\newcommand{\Q}{\mbox{$\cal Q$}}

\newcommand{\flycheck}{{\sf flycheck}\xspace}
\newcommand{\ciaopp}{{\sf CiaoPP}\xspace}
\newcommand{\ciao}{{\sf Ciao}\xspace}

\newcommand{\verifly}{{\sf VeriFly}\xspace}

\newcommand{\hcir}{HC IR\xspace}

\newcommand{\lsem}{\mbox{$\llbracket$}}
\newcommand{\rsem}{\mbox{$\rrbracket$}}
\newcommand{\sem}[1]{\lsem #1 \rsem}

\newcommand{\prog}{P}
\newcommand{\psem}{\sem{\prog}_{\Q}}

\newcommand{\analysis}{\ensuremath{\psem^\alpha}}
\newcommand{\renanalysis}[1]{\ensuremath{{\it Ren}(#1, \analysis)}}

\newcommand{\underapp}[1]{#1^{\alpha-}}
\newcommand{\overapp}[1]{#1^{\alpha+}}

\newcommand\vv{{\bar v}}

\newcommand\pv{p(\bar v)}
\newcommand\pvp{p(\bar v')}

\newcommand{\secbeg}{\vspace*{-2mm}}
\renewcommand{\secbeg}{}
\newcommand{\secend}{\vspace*{-2mm}}
\renewcommand{\secend}{}
\newcommand{\parbeg}{\vspace*{-4mm}}
\renewcommand{\parbeg}{}

\newcommand{\figbeg}{\vspace*{-3mm}}
\renewcommand{\figbeg}{}
\newcommand{\figend}{\vspace*{-3mm}}

\title[\verifly: On-the-fly Assertion Checking via Incrementality]{
\verifly: On-the-fly Assertion Checking\\ via Incrementality$^{
    \thanks{%
      Research partially funded by MINECO MICINN PID2019-108528RB-C21
      \emph{ProCode} project, FPU grant 16/04811, the Madrid
      M141047003 \emph{N-GREENS} and P2018/TCS-4339 \emph{BLOQUES-CM}
      programs, and the Tezos foundation. We are also grateful to the
      anonymous reviewers for their comments.}}$ }

\author[M. A. Sanchez-Ordaz et al.]{
  \hspace*{-6mm}MIGUEL A. SANCHEZ-ORDAZ$^{1,2}$,
  ISABEL GARCIA-CONTRERAS$^{1,2}$,
  VICTOR PEREZ$^{1,2}$,\authorbreak
  \hspace*{-13mm}JOS\'{E} F.~MORALES$^{1,2}$,
  PEDRO LOPEZ-GARCIA$^{1,3}$ and
  MANUEL V.~HERMENEGILDO$^{1,2}$\\
  \hspace*{-13mm}$^1$IMDEA Software Institute,\\
  \hspace*{-13mm}$^2$Universidad Polit\'{e}cnica de Madrid (UPM),\\
  \hspace*{-13mm}$^3$Spanish Council for Scientific Research (CSIC)\\
  \hspace*{-13mm}\email{\{ma.sanchez.ordaz,isabel.garcia,victor.perez,josef.morales,pedro.lopez,manuel.hermenegildo\}@imdea.org}
}

\begin{document}
\maketitle

\begin{abstract}

  Assertion checking is an invaluable programmer's tool for finding
  many classes of errors or verifying their absence in dynamic
  languages such as Prolog.
  For Prolog programmers this means being able to have relevant
  properties such as modes, types, determinacy, non-failure, sharing,
  constraints, cost, etc., checked and errors flagged without having
  to actually run the program.
  Such global static analysis tools are arguably most useful the
  earlier they are used in the software
  development cycle, and fast response times
  are 
  essential
  for
  interactive use.
  Triggering a full
  and precise
  semantic analysis of a software project every time a change is made can be
  prohibitively expensive. This is specially the case when
  complex properties need to be
  inferred 
  for 
  large, realistic code bases.
  In 
  our static analysis and verification framework this challenge is
  addressed through a combination of modular and incremental (context-
  and path-sensitive) analysis 
  that is responsive to program edits, at different levels of
  granularity.
  In this tool paper we present 
  how the combination of
  this framework within an integrated development environment (IDE) 
  takes advantage of such incrementality to achieve a high level of reactivity
  when reflecting analysis and verification results back as colorings and
  tooltips directly on the program text---the tool's \verifly mode. The
  concrete implementation that we describe is Emacs-based and
  reuses in part 
  off-the-shelf ``on-the-fly'' \emph{syntax} checking facilities
  (\flycheck).
  We believe that similar extensions are also reproducible with low
  effort in other mature development environments.
  Our initial experience with the tool shows quite promising results,
  with low latency times that provide early, continuous, and precise
  assertion checking and other semantic feedback to programmers during
  the development process.
  The tool supports Prolog natively, as well as other languages by
  semantic transformation into Horn clauses.
  This paper is under consideration for acceptance in
  TPLP.
\end{abstract}

\secbeg
\section{Introduction}
\secend

Global static analysis and verification tools can greatly help
developers detect high-level, semantic errors in programs or certify
their absence. This is specially the case for high-level, dynamic
languages such as Prolog, where obtaining information such as modes,
types, determinacy, non-failure, cardinality, variable sharing in
structures, constraints, termination, cost, etc., and checking it
against specifications, in the form of annotations or assertions, can
be invaluable aids to programmers. 

Arguably, such tools are more effective the earlier the
stage in which they are applied within the software development
process. Particularly useful is their application
during
code development, simultaneously with the code writing process,
alongside the compiler, debugger, etc.
The tight integration of global analysis and verification
at such early stages
requires fast response times and
source-level presentation of the results within the code development
environment, in order to provide timely and useful feedback to the
programmer.
However, triggering a full and precise semantic analysis of a software
project every time a change is made can be expensive and may not be
able to meet the requirements of the scenario described. This is
specially the case when complex properties need to be inferred for
large, realistic code bases.

Our approach builds on the \ciaopp program development
framework~\cite{aadebug97-informal-short,prog-glob-an-short,assrt-theoret-framework-lopstr99-short,ciaopp-sas03-journal-scp-short,intermod-incanal-2020-tplp-short},
which performs combined static and dynamic program analysis, assertion
checking, and program transformations, based on computing provably
safe approximations of properties, generally using the technique of
abstract interpretation~\cite{Cousot77}.
This framework supports natively Prolog and several extensions within
the LP and CLP paradigms, and can also be applied to
other high- and low-level languages, using the now 
well-understood technique of semantic translation into intermediate
Horn clause-based representation. 
The framework has many uses, but one of the main ones is precisely as
an aid for the programmer during program development, since it can
capture semantic errors that are significantly higher-level than those
detected by classical compilers, as well as produce certificates that
programs do not violate their assertions, eliminate run-time
assertion tests, etc. 

In our framework, 
the requirements for fast response time and precision
stemming from interactive use pointed out above are
addressed 
through a number of techniques, and in particular by an efficient
fixpoint engine, which 
performs context- and path-sensitive
inter-procedural analysis \emph{incrementally},
i.e.,
reactively to fine-grain
edits, avoiding
reanalyses where possible, both within modules and across the
modular organization of the code into
separate compilation units.
In this tool paper we illustrate how the integration of
the static analysis and verification framework within an integrated
development environment (IDE) 
takes advantage of the incremental and modular features
to achieve a high level of reactivity when reflecting analysis and
verification results back as colorings and tooltips directly on the
program text---the tool's \verifly mode.
The concrete integration 
described
builds on an existing
Emacs-based development environment for the \ciao language, 
and reuses in part
off-the-shelf ``on-the-fly''
\emph{syntax} checking capabilities offered by the Emacs \flycheck
package.
Emacs was chosen because it is a solid platform and preferred by many
experienced users.
However, this low-maintenance approach should be easily reproducible in other
modern extensible editors and mature program development environments.

\secbeg
\section{The \ciaopp Framework}\label{sec:ciaopp-framework}
\secend

\begin{figure}[t]
  \figbeg
  \centering
  \resizebox{1\textwidth}{!}{
\pgfdeclarelayer{background}
\pgfdeclarelayer{foreground}
\pgfsetlayers{background,main,foreground}

\newcommand{\tikzmark}[1]{\tikz[overlay,remember picture] \node (#1) {};}
\tikzset{square arrow/.style={to path={-- ++(0,.25) -| (\tikztotarget)}}}

\tikzstyle{source}=[draw, draw=cyan!80!black!100, fill=cyan!20, text width=6em, font=\sffamily,
    thick,
    minimum height=2.5em,drop shadow]
\tikzstyle{transform}=[draw, draw=cyan!80!black!100, fill=cyan!20, text width=5em, font=\sffamily,
thick,  densely dashed,
    minimum height=2.5em,drop shadow]
\tikzstyle{file}=[draw, draw=Peach!80!black!100, fill=Peach!20, text width=0.8cm, font=\sffamily,
    thick,
    minimum height=2.5em,drop shadow]
\tikzstyle{emacs}=[draw, draw=Plum!80!black!100, fill=Plum!20, text width=6em, font=\sffamily,
    thick,
    minimum height=2.5em,drop shadow]
\tikzstyle{tool}=[draw=green!50!black!100, fill=green!40, text width=5em, font=\sffamily, 
    thick,
    text centered, 
    chamfered rectangle, chamfered rectangle angle=30, chamfered rectangle
    xsep=2cm]
\tikzstyle{ciaopp-db}=[draw, draw=Violet!80!black!100, fill=Violet!20, text width=6.2em, font=\sffamily,
    thick,
    minimum height=2.5em,drop shadow]
\tikzstyle{source-db}=[draw, draw=Violet!80!black!100, fill=Violet!20, text width=5em, font=\sffamily,
    thick,
    minimum height=2.5em,drop shadow]
\tikzstyle{newtool}=[draw=yellow!50!black!100, fill=yellow!40, text width=5em, font=\sffamily, 
    thick,
    text centered, 
    chamfered rectangle, chamfered rectangle angle=30, chamfered rectangle xsep=2cm]
\tikzstyle{midresult}=[draw, fill=white!40, text width=5em, font=\sffamily,
    thick,
    rounded rectangle,
    minimum height=1em,drop shadow]
\tikzstyle{warnresult}=[color=orange!50!black!100, fill=orange!40, text width=4em, font=\sffamily, 
    thick,
    rounded corners=2pt,
    minimum height=1em,drop shadow]
\tikzstyle{errresult}=[color=red!80!black!100, fill=red!20, text width=4em, font=\sffamily, 
    thick,
    rounded corners=2pt,
    minimum height=1em,drop shadow]
\tikzstyle{okresult}=[color=green!50!black!100, fill=green!40, text width=4em, font=\sffamily, 
    thick,
    rounded corners=2pt,
    minimum height=1em,drop shadow]

\scriptsize
\begin{tikzpicture}[>=latex]
  

  
  \node (src1) [file] {src v1};
  \node (src2) [file,below of=src1] {src v2};
  \node (src3) [file,below of=src2] {src v3};

  \node (code) [transform, right of=src2, xshift=10mm] {
    \textbf{Transform}
  };

  \node (src-db) [source-db, below of=code] {
    \textbf{Source DB}
  };




  \draw [thick, ->] (src1.east) to (src-db.west);
  \draw [thick, ->] (src2.east) -- node [yshift=2.5mm,rotate=-32]
    {}
    (src-db.west);
  \draw [thick, ->] (src3.east) -- node [yshift=-2mm]
    {}
    (src-db.west);

  \draw [thick, ->] (src-db) -- (code);

  
  \path (code)+(11.2em,2em) node (code-db) [ciaopp-db] {\textbf{Clause DB}};

\node (lib-db) [ciaopp-db,below of=code-db,yshift=3mm] {\textbf{Libraries DB}};
\node (normalizer) [below of=lib-db,ciaopp-db] {\textbf{Assertion DB}};
\node (statana) [above of=code-db,yshift=1em,tool] {Static Analyzer};

\node (anainfo) [ciaopp-db,right of=statana,xshift=6em] {\textbf{Analysis DB}};
\node (ctchecker) [tool,right of=normalizer,xshift=6em] {Static Checker};
\path [draw, thick, ->] (anainfo) -- node [] {} (ctchecker) ;
\node (true)      [color=green!50!black,above of=ctchecker,midresult] {:- true};

\path[color=blue] (ctchecker)+(8.5em,2em) node (check) [midresult] {:- check};
\path[color=red] (ctchecker)+(8.5em,-0em) node (false) [midresult] {:- false};
\path[color=green!50!black] (ctchecker)+(8.5em,-2em) node (checked) [midresult] {:- checked};
\node (spec) [tool,above of=check,yshift=1mm] {Dynamic Annotator};
\path (ctchecker)+(16.5em,0em) node (cterror) [errresult] {Error};
\path (ctchecker)+(16.5em,2em) node (verifwarn) [warnresult] {Warning};
\path (ctchecker)+(16.5em,-2em) node (verified) [okresult] {Verified};
\node (verified-side) [right of=verified,xshift=2mm] {};
\node (srct) [file,right of=spec,xshift=4em] {RT safe src};

\path [draw, thick, ->] (code-db.north) -- node [xshift=1em] {{\sf ($\Delta$)}} (statana.south) ;
\path [draw, thick, ->] (code-db.east) -- node [] {} (spec) ;
\path [draw, thick, <->] (code.east) -- node [rotate=20,xshift=0.1em,yshift=0.8em] {{\sf $\Delta$ CHC}} (code-db.west) ;
\path [draw, thick, ->] (code.east) -- node [] {} (normalizer.west) ;
\draw [thick, ->] (lib-db.east) to [bend right] (statana.south east) ;
\path [draw, thick, ->] (normalizer) -- node [] {} (ctchecker) ;
\path [draw, thick, <->] (statana) -- node [] {} (anainfo.west) ;
\path [draw, thick, ->] (ctchecker.east) -- node [] {} (false.west) ;
\path [draw, thick, ->] (ctchecker.east) -- node [] {} (check.west) ;
\path [draw, thick, ->] (ctchecker.east) -- node [] {} (checked.west) ;
\path [draw, thick, ->] (false.east) -- node [] {} (cterror.west) ;
\path [draw, thick, ->] (check.east) -- node [] {} (verifwarn.west) ;
\path [draw, thick, ->] (checked.east) -- node [] {} (verified.west) ;

\draw [draw, thick, ->] (check.north) to (spec.south) ;
\path [draw, thick, ->] (spec) -- node [xshift=1em] {} (srct) ;

\path [color=black] (statana.north)+(0,1.5em) node (preprocessor) {\sffamily \small \textbf{CiaoPP}};
\path [color=black] (code.north)+(0,1.5em) node (program) {\sffamily \small Front-end};

\begin{pgfonlayer}{background}
  \path (statana.north west)+(-3.3,0.4) node (g) {};
  \path (checked.south east)+(0.5,-0.2) node (h) {};
  
  \path[fill=yellow!20,rounded corners, draw=black!50, densely dashed] (g) rectangle (h);
\end{pgfonlayer}

\begin{pgfonlayer}{background}
  \path (code.north west)+(-0.1,0.8) node (g) {};
  \path (src-db.south east)+(0.1,-0.1) node (h) {};
  
  \path[source] (g) rectangle (h);
\end{pgfonlayer}

  

  

\end{tikzpicture}

  }
  \figend  
  \figend  
  \caption{Architecture of the \ciaopp framework.}\label{fig:ciaopp}
  \figend
  \vspace*{-1mm}
\end{figure}
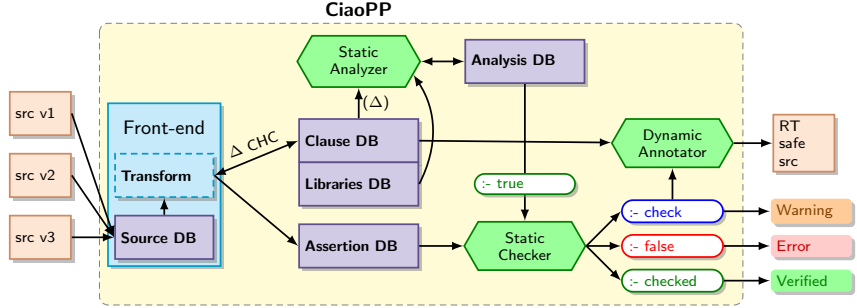

We start by providing an informal overview of the components and
operation of the 
\ciaopp framework
(Fig.~\ref{fig:ciaopp}).

\parbeg
\paragraph{\textbf{Front end.}}

Before getting into the analysis and verification phases, the tool's
front end 
transforms any syntactic or semantic extensions used in the input
program (such as, e.g, functional notation, or, more specifically for
our discussion, the syntactic sugar related to assertions) in order to
reduce each module to the intermediate representation that the
framework works with.  This representation, which we refer to as the
\hcir, is fundamentally plain Horn clauses (plain Prolog) extended
with the (canonical part) of the assertion language, reviewed in the
next section.
Although beyond the scope of this paper, as mentioned in the
introduction, the framework can also be applied to other input
languages, outside (C)LP, which is done by translating such input
programs to the same
\hcir~\cite{decomp-oo-prolog-lopstr07-short}.\footnote{
  This approach is used nowadays in many analysis and
  verification tools
~\cite{
  Peralta-Gallagher-Saglam-SAS98,
  HGScam06-short,decomp-oo-prolog-lopstr07-short,
  mod-decomp-jist09-short,DBLP:conf/tacas/GrebenshchikovGLPR12-short,
  DBLP:conf/cav/GurfinkelKKN15-shorter,
  AngelisFPP17,
  resource-verification-tplp18-shortest,
  resources-blockchain-sas20-short,
  big-small-step-vpt2020-shorter}.
The front end is also in charge of these translations, as well as of
translating the analysis and verification results back to the source
language.
Techniques such as partial evaluation and program specialization offer
powerful methods to obtain such translations with provable
correctness---see~\cite{anal-peval-horn-verif-2021-tplp} for a recent survey.}
However, herein we concentrate on the native use of the tool for (C)LP
programs.  The examples will be written in
\ciao~Prolog~\cite{hermenegildo11:ciao-design-tplp,ciao-reference-manual-1.20-short}\footnote{\url{https://github.com/ciao-lang/devenv}}
using its assertions library and some additional syntactic sugar such
as functional notation, but the tool can easily be adapted to process
source files for other Prolog flavors and systems.
The framework itself is also written in \ciao.

\parbeg
\paragraph{\textbf{The assertion language.}}

A fundamental element of the framework 
is its \emph{assertion
  language}~\cite{full-prolog-esop96-short,prog-glob-an-short,assert-lang-disciplbook-short}.
Such assertions can express a wide range of properties, including
functional (state) properties, such as modes, types,
sharing/aliasing, constraints, etc.,
as well as non-functional (i.e., global, computational)
properties, such as determinacy, non-failure, cardinality, or
resource usage (energy, time, memory, \ldots). The set of properties is
extensible and new abstract domains (see the later discussion of the
analysis) can be defined as ``plug-ins'' to support them.  Assertions
associate these properties to different program points, and
are used 
for multiple purposes,
including writing specifications, 
reporting static analysis and verification results (to the programmer, other
parts of the framework, or other tools), providing assumptions, describing
unknown code, generating test cases automatically, or producing documentation.

\noindent
{\sc Predicate-level assertions}
allow describing \emph{sets} of \emph{preconditions} and \emph{conditional
  postconditions} on the state for a given 
predicate (as well as global properties).
In particular, \pred assertions are of the form:\\ [-2em]
\begin{center}
  \hltexttt{:- [ \mbox{\status} ] pred \mbox{\head} [: \mbox{\pre} ] [=>
    \mbox{\post} ] [+ \mbox{\comprops} ].}
\end{center}
\vspace*{-1mm}
\noindent
where \head is a predicate descriptor
that denotes the predicate that the assertion applies to, and \pre and
\post are conjunctions of \emph{state property} literals (the modes,
types, constraints, etc.), which can include arguments, and these can
also be variables from \head.
Such properties are predicates, typically written in
(subsets of) 
the source
language, and thus runnable, so that
they can be used as run-time checks, and which (see ``Assertion
verification'' later in this section) may be
abstracted and inferred by some domain in \ciaopp.
They can be imported from system libraries or user-defined.
\pre expresses properties that hold when \head is called.  \post
states properties that hold if \head is called in a state compatible
with \pre and the call succeeds. Both \pre and \post can be empty
conjunctions (meaning true), and in that case they can be omitted.
\comprops describes properties of the whole computation such as
determinism, non-failure, resource usage,
termination, etc., and they also apply
to calls to the predicate 
that meet \pre.
Assertions are optional, and there can be multiple \texttt{pred}
assertions for a predicate.
This allows describing different behaviors for the same predicate for
different call substitutions, i.e., different \emph{modes}.
If multiple \texttt{pred} assertions are present for a given
predicate, 
their \pre fields then describe the set of all admissible calls to the
predicate, i.e., at least one \pre must hold for each call to \head.

$\status$
is a qualifier of the meaning of the assertion. Here we consider (in
the context of static assertion
checking)
the following $\status$es (cf.\ Fig.~\ref{fig:ciaopp}):

\vspace*{-2mm}
\begin{itemize}
  \itemsep=0pt
\item {\tt \acheck}: the assertion expresses properties that must hold at
  run-time, i.e., that the analyzer should prove (or else generate run-time
  checks for). {\tt check} is the \emph{default} status, 
  and can be omitted.
\item {\tt \achecked}: the analyzer proved that the property holds in all executions.
\item {\tt \afalse}: the analyzer proved that the property does not hold in some execution.
\end{itemize}

Additionally, assertions with \texttt{trust} status can be used to
provide the analyzer with information that it should \emph{assume}. This
can be useful for, e.g., describing external or unknown code or 
improving analysis precision.

Fig.~\ref{fig:nrevf} 
provides some examples of assertions. 
The syntax that allows including the \pred assertions is made
available through the \texttt{assertions} package (packages are
similar to modules but oriented towards providing syntactic and semantic
extensions).
Predicates \texttt{nrev/2} and \texttt{conc/2} define naive reverse
and are written using the \texttt{functional} notation package.
The first assertion (line~\ref{asrtone})
expresses that calls to \texttt{nrev/2} with the first argument bound
to a list are admissible, and that if such calls succeed, then the
second argument should also be bound to a list (as well as being
non-failing, deterministic, terminating, and quadratic). 
Thus, such an assertion
provides information on (a generalization of) types and modes, as well
as determinism, failure, cost, etc.

\begin{figure}[t]
\figbeg
  \prettylstciao
\begin{lstlisting}[belowskip=0mm,escapechar=@]
:- module(_, [nrev/2], [assertions, nativeprops, functional]).
:- entry nrev/2 : {list, ground} * var.                 @\label{entry}@
:- use_module(someprops).

:- pred nrev(A, B) : list(A) => list(B) 
   + (not_fails,is_det,terminates,steps_o(exp(length(A),2)). @\label{asrtone}@
nrev([])    := [].
nrev([H|L]) := ~conc(nrev(L),[H]).

:- pred conc(A,B,C) : list(A) => size_ub(C,length(A)+length(B)) + steps_o(length(A)).
conc([],    L) := L.
conc([H|L], K) := [ H | conc(L,K) ]. 
\end{lstlisting}
\figend
\caption{Naive reverse with some assertions.}\label{fig:nrevf}
\figend
\end{figure}

\begin{figure}[t]
\figbeg
  \prettylstciao
\begin{lstlisting}[belowskip=0mm,escapechar=]
:- module(someprops, _, [functional, hiord, assertions]).

:- prop color/1.    
color := red | blue | green. % Equivalent to: color(red). color(blue). color(green).

:- prop list/1.     
list := [] | [_ | list].     % Equivalent to: list([]). list([_|T]) :- list(T).

:- prop list/2.  
list(T) := [] | [~T | list(T)].

:- prop sorted/1.   
sorted := [] | [_].
sorted([X,Y|Z]) :- X @< Y, sorted([Y|Z]).
\end{lstlisting}
\figend
\caption{Examples of state \texttt{prop}erty definitions.}\label{fig:fmoreprops}
\figend
\end{figure}

Fig.~\ref{fig:fmoreprops} provides some examples of state properties
(\texttt{prop}s, for short), which, as mentioned before, 
are defined and exported/imported as normal
predicates: e.g., in Fig.~\ref{fig:nrevf} \texttt{list/1},
\texttt{list/2}, or \texttt{color/1} are imported from the 
\texttt{someprops} user module 
while others (e.g., \texttt{size\_ub/2}) from the system's
\texttt{nativeprops} library.
Properties need to be marked explicitly as such
with a \texttt{:- prop} declaration, 
and this flags that they need to meet some
restrictions~\cite{assert-lang-disciplbook-short,ciao-reference-manual-1.20-short}.
E.g., their execution should terminate 
for any possible call since, as 
discussed later, 
\texttt{prop}s will not only be checked at compile time, 
but may also be involved in run-time checks. 
Types are just a particular case (further restriction) of state
properties.  Different type systems
are provided as libraries.
Properties like \texttt{list/1} 
that are in addition regular
types can be flagged as such: \verb|:- prop list/1 + regtype.|
or, more compactly, 
\verb|:- regtype list/2.|

The combination of multiple assertions for a given predicate is
defined more precisely as follows: 
given a
set of assertions $\{a_1, \ldots, a_n\}$, with $a_i = ``\texttt{:- pred }$ $\head
  \texttt{ : } \pre_i \texttt{ => } \post_i \texttt{.}$'' the set of
  \emph{assertion conditions} for \head is $\{ C_0, C_1, \ldots , C_n\}$,
  with:
  \[
  \begin{array}{c}
  \ C_0 = \callsAsr{\head}{\vee_{j = 1}^{n} \pre_j}, \ and \\
  C_i = \successAsr{\head}{\pre_i}{\post_i}, i = 1\ldots n
  \end{array}
  \]
where $\callsAsr{\head}{\pre}$
states conditions on all concrete calls to the predicate described by
\head, and $\successAsr{\head}{\pre_i}{\post_i}$ the conditions
on the success constraints produced by calls to \head if $\pre_i$ is
satisfied.
If the assertions $a_i$ above, $i=1, \ldots, n$, include a
\hltexttt{\mbox{+ \comprops}} field, then the set of \emph{assertion
  conditions} also includes conditions of the form:
$\compAsr{\head}{\pre_i}{\comprops_i}$, for $i=1, \ldots, n$, that express
properties of the whole computation for calls to \head if $\pre_i$ is
satisfied.

\medskip
\noindent
In addition to predicate-level assertions, {\sc program-point
  assertions} are of the form:
\vspace*{-1mm}
  \prettylstciao
\begin{lstlisting}[belowskip=0mm,escapechar=@,frame=,numbers=none]
      @\emph{Status}@(@\emph{StateFormula}@),
\end{lstlisting}
  and they can be placed at program locations where a new literal may
  be added.  E.g., for status {\tt \acheck} they should be interpreted
  as ``whenever computation reaches a state corresponding to this
  program point,
\emph{StateFormula} should hold.'' For example,
\noindent
  \prettylstciao
\begin{lstlisting}[belowskip=0mm,escapechar=@,frame=,numbers=none]
      check((list(color, A), var(B))), 
\end{lstlisting}
  \noindent is a program-point assertion 
that checks whether \texttt{A} is instantiated to a list of colors and
\texttt{B} is a variable, where \texttt{A} and
\texttt{B} are variables of the clause where the assertion appears.

\parbeg
\paragraph{\textbf{Assertion-related syntactic sugar.}}

In order to facilitate writing assertions, as well as compatibility
with different systems, the assertion language also provides syntactic
sugar such as \emph{modes}, Cartesian product notation, markdown
syntax, etc.
For example, 
the following set of \pred assertions use Cartesian product
notation to provide information on a reversible sorting predicate:
\noindent \prettylstciao
\vspace*{-1mm}
\begin{lstlisting}[belowskip=0mm,escapechar=@,frame=,numbers=none]
:- pred sort/2 : list(num) * var => list(num) * list(num) + is_det. 
:- pred sort/2 : var * list(num) => list(num) * list(num) + non_det. 
\end{lstlisting}
\noindent (in addition, curly brackets can be used to group
properties---see Fig.~\ref{fig:nrevf}). 
The assertion language also allows defining \emph{modes}, which are
macros that can be placed in argument positions of \head and expand to
properties in different assertion fields.  For example, using
the \texttt{isomodes} library 
the assertions above can also be expressed as:
\vspace*{-1mm}
\begin{lstlisting}[escapechar=@,frame=,numbers=none]
:- pred sort(+list(num), -list(num)) + is_det.
:- pred sort(-list(num), +list(num)) + non_det.
\end{lstlisting}
\vspace*{-1mm}
Or, if no types and only modes are used:
\vspace*{-1mm}
\begin{lstlisting}[escapechar=@,frame=,numbers=none]
:- pred sort(+, -).
:- pred sort(-, +).
\end{lstlisting}
\vspace*{-1mm}
There is also an alternative encoding via ``markdown-style'' comments
(provided by libraries \texttt{doccomments} and \texttt{markdown})
that allows writing, e.g.:
\vspace*{-1mm}
\begin{lstlisting}[basicstyle=\ttfamily\small,escapechar=@,frame=,numbers=none]
%! sort(+list(num),-list(num)): This predicate sorts.
\end{lstlisting}
\vspace*{-1mm}
Thanks to these libraries and the underlying syntactic expansion
mechanisms, and their modular nature, it is easy to adapt the tool to
other syntactic forms, on a per-module basis, such as for example
supporting ``Quintus-style'' modes:
\noindent
\vspace*{-1mm}
\begin{lstlisting}[escapechar=@,frame=,numbers=none]
:- mode sort(+, -).
:- mode sort(-, +).
\end{lstlisting}
\vspace*{-1mm}
or the SWI-Prolog \texttt{pldoc}-style annotations for documentation,
making all of them machine checkable.

\parbeg
\paragraph{\textbf{Static Program Analysis:}}
The \ciaopp~verification framework uses analyses based on
abstract interpretation (the ``Static Analyzer'' in
Fig.~\ref{fig:ciaopp})
to compute safe over-approximations of the program
semantics. 
Given a program $P$ and a set of queries $\Q$, an analysis graph is inferred
that abstracts the (possibly infinite) set of (possibly infinite) execution trees.
The analysis result is denoted by $\psem^\alpha$, where $\alpha$ is the abstraction(s)
performed. By default, the abstract domains to be used are
automatically selected, depending the properties that appear in the
assertions, but manual selection is also possible. 
Nodes in this
graph abstract how predicates are called (i.e., how they are used in
the program). A predicate may have several nodes if there are
different calling situations (also known as context-sensitivity). For
each calling situation, properties that hold if the predicate succeeds
are also inferred,
and can be
represented by ({\tt \atrue}) assertions.
For our purposes, and without loss of generality, we treat $\analysis$ as a set
of tuples (one per node): 
$\{\tuple{L_1,\lambda_{1}^c,\lambda_{1}^s}, \ldots,
\tuple{L_n,\lambda_{n}^c,\lambda_{n}^s}\}$. In each
$\tuple{L_i,\lambda_{i}^c,\lambda_{i}^s}$ triple, $L_i$ is
a predicate descriptor, and $\lambda_{i}^c$ and $\lambda_{i}^s$ are,
respectively, the abstract call and success substitutions, elements of
abstract domain $D_\alpha$.
\footnote{As mentioned
  before, abstract domains are defined as plug-ins which provide the
  basic abstract domain lattice operations and transfer functions, and
  are made accessible to the generic fixpoint computation component.}
The edges in the graph capture how predicates call each other. Hence this
analysis graph also provides an abstraction of the paths explored by the
concrete executions through the program (also known as path-sensitivity).
The analysis graph thus embodies two different abstractions (two
different abstract domains): the graph itself is a \emph{regular
  approximation} of the paths through the program, using a domain of
regular structures. Separately, the abstract values (call and success
patterns) contained in the graph nodes are finite representations of
the states occurring at each point in these program paths, by means of
one or more data-related abstract domains. 

\parbeg
\paragraph{\textbf{Assertion verification:}}
To verify a program, the behaviors abstracted in $\psem^\alpha$ are
compared 
with
program assertions (the ``Static Checker'' in Fig.~\ref{fig:ciaopp}).
The verification results are reported as changes in the status and
transformations of the assertions: {\tt \achecked}, if the properties
are satisfied; {\tt \afalse} if some property was proved not to hold;
or {\tt \acheck} if it was not possible to determine any the first
two, in which case run-time checks will be included in the program to
make it run-time safe (Fig.~\ref{fig:ciaopp}).
Since in our framework the intended semantics is also specified
by using predicates, some of the properties may be
undecidable and also not exactly representable in the abstract
domains.
However, it is possible to under- and over-approximate them, denoted by
$\underapp{I}$ and $\overapp{I}$.
Such approximations are always computable by choosing the closest
element in the abstract domain. At the limit $\bot$ and $\top$ are,
respectively, an under-approximation and an over-approximation of any
specification.
Let $\renanalysis{\pv}$, where $\pv$ is a predicate descriptor, denote
the set of tuples resulting from the static analysis for predicate
$p\ in \ \prog$ w.r.t.\ $\Q$ renamed to variables $\vv$, i.e.:
$\renanalysis{\pv} = \{\tuple{\pv,
  \lambda^{c}\sigma,\lambda^{s}\sigma} \mid \exists
\tuple{\pvp,\lambda^c,\lambda^s} \in \analysis
\text{ and } \sigma
\text{ is a renaming }$ \\ $\text{substitution s.t. }
\pv= \pvp\sigma\}$.

\noindent
The following are some sufficient conditions, adapted
from~\citeauthor{assrt-theoret-framework-lopstr99-short}~(\citeyear{assrt-theoret-framework-lopstr99-short}),
and~\citeauthor{mod-ctchecks-lpar06-short}~(\citeyear{mod-ctchecks-lpar06-short}),
which allow deciding whether assertions are \texttt{\achecked} or
\texttt{\afalse} with respect to the concrete semantics, based on the
abstract semantics:

\begin{itemize}

\item $\callsAsr{\pv}{\pre}$ is \emph{checked} (resp. \emph{false})
  for predicate $p \in \prog$ w.r.t.\ $\Q$ if \\ $\forall
  \tuple{\pv,\lambda^c,\lambda^s} \in \renanalysis{\pv}:
  \ \lambda^c \sqsubseteq \underapp{\pre}$
  (resp. $\lambda^c \sqcap \overapp{\pre} = \bot$).
  
\item $\successAsr{\pv}{\pre}{\post}$ is \emph{checked}
  (resp. \emph{false}) for predicate $p \in P$ w.r.t.\ $\Q$ if
  $\forall \tuple{\pv, \lambda^c,\lambda^s} \in \renanalysis{\pv}:
\ ( (\lambda^c
\sqcap \overapp{\pre} = \bot) \lor (\lambda^s \sqsubseteq
\underapp{\post}))$ \\
(resp. $\lambda^c \sqsubseteq \underapp{\pre} \land (\lambda^s \sqcap
\overapp{\post} = \bot)$ and there is at least one actual call to
predicate $p$ that succeeds).

\end{itemize}

\parbeg
\paragraph{\textbf{Supporting incrementality.}}
In order to support incrementality, the analysis graph produced by the
static analyzer is made persistent (``Analysis DB'' in
Fig.~\ref{fig:ciaopp}), storing
an abstraction of the behavior of each predicate and predicate (abstract) call
dependencies. In turn, the ``Front-end'' (Fig.~\ref{fig:ciaopp})
keeps track of and translates source code changes into, e.g., clause
and assertion additions, deletions, and changes in the intermediate
representation ($\mathsf{\Delta}$~{\sf CHC} in the figure).
These changes are transmitted to the static analyzer, which performs
incremental fixpoint computation. This process consists in finding
the parts of the graph that need to be deleted or recomputed,
following their dependencies, and updating the fixpoint
~\cite{intermod-incanal-2020-tplp-short,incanal-assrts-openpreds-lopstr19-post-short,inc-fixp-sas-short,incanal-toplas-short}.
The key point here is that a tight relation between the analysis results and the
predicates in the program is kept, allowing reducing the re-computation to the
part of the analysis that corresponds to the affected predicates, and only
propagating it to the rest of the analysis graph if necessary.

\secbeg
\section{\verifly: The On-the-fly IDE Integration}\label{sec:verifly}
\secend

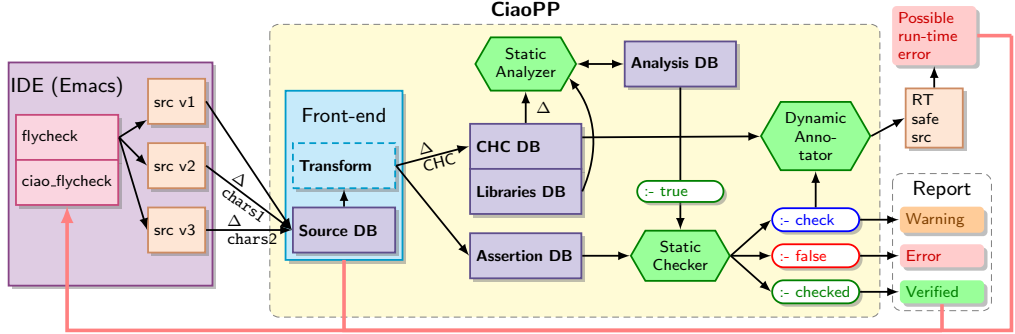
\begin{figure}
  \hspace*{-2cm}
 \resizebox{1.15\textwidth}{!}{
\pgfdeclarelayer{background}
\pgfdeclarelayer{foreground}
\pgfsetlayers{background,main,foreground}

\newcommand{\tikzmark}[1]{\tikz[overlay,remember picture] \node (#1) {};}
\tikzset{square arrow/.style={to path={-- ++(0,.25) -| (\tikztotarget)}}}

\tikzstyle{source}=[draw, draw=cyan!80!black!100, fill=cyan!20, text width=6em, font=\sffamily,
    thick,
    minimum height=2.5em,drop shadow]
\tikzstyle{transform}=[draw, draw=cyan!80!black!100, fill=cyan!20, text width=5em, font=\sffamily,
thick,  densely dashed,
    minimum height=2.5em,drop shadow]
\tikzstyle{file}=[draw, draw=Peach!80!black!100, fill=Peach!20, text width=0.7cm, font=\sffamily,
    thick,
    minimum height=2.5em,drop shadow]
\tikzstyle{flycheck}=[draw, draw=RubineRed!80!black!100, fill=RubineRed!20, text width=5em, font=\sffamily,
    thick,
    minimum height=2.5em,drop shadow]
\tikzstyle{emacs}=[draw, draw=Plum!80!black!100, fill=Plum!20, text width=6em, font=\sffamily,
    thick,
    minimum height=2.5em,drop shadow]
\tikzstyle{tool}=[draw=green!50!black!100, fill=green!40, text width=3.5em, font=\sffamily, 
    thick,
    text centered, 
    chamfered rectangle, chamfered rectangle angle=30, chamfered rectangle
    xsep=2cm]
\tikzstyle{ciaopp-db}=[draw, draw=Violet!80!black!100, fill=Violet!20, text width=5.5em, font=\sffamily,
    thick,
    minimum height=2.5em,drop shadow]
\tikzstyle{source-db}=[draw, draw=Violet!80!black!100, fill=Violet!20, text width=5em, font=\sffamily,
    thick,
    minimum height=2.5em,drop shadow]
\tikzstyle{newtool}=[draw=yellow!50!black!100, fill=yellow!40, text width=5em, font=\sffamily, 
    thick,
    text centered, 
    chamfered rectangle, chamfered rectangle angle=30, chamfered rectangle xsep=2cm]
\tikzstyle{midresult}=[draw, fill=white!40, text width=4em, font=\sffamily,
    thick,
    rounded rectangle,
    minimum height=1em,drop shadow]
\tikzstyle{warnresult}=[color=orange!50!black!100, fill=orange!40, text width=4em, font=\sffamily, 
    thick,
    rounded corners=2pt,
    minimum height=1em,drop shadow]
\tikzstyle{errresult}=[color=red!80!black!100, fill=red!20, text width=4em, font=\sffamily, 
    thick,
    rounded corners=2pt,
    minimum height=1em,drop shadow]
\tikzstyle{okresult}=[color=green!50!black!100, fill=green!40, text width=4em, font=\sffamily, 
    thick,
    rounded corners=2pt,
    minimum height=1em,drop shadow]

\scriptsize
\begin{tikzpicture}[>=latex]
  

  \node (flycheck) [flycheck]  {flycheck};
  \node (ciao-flycheck) [flycheck,below of=flycheck,yshift=3mm] {ciao\_flycheck};
  
  \node (src1) [file,right of=flycheck,yshift=2em,xshift=0.7cm] {src v1};
  \node (src2) [file,below of=src1] {src v2};
  \node (src3) [file,below of=src2] {src v3};

  \node (code) [transform, right of=src2, xshift=16mm] {
    \textbf{Transform}
  };

  \node (src-db) [source-db, below of=code] {
    \textbf{Source DB}
  };



  \draw [thick, ->] (flycheck.east) -- (src1);
  \draw [thick, ->] (flycheck.east) -- (src2.west);
  \draw [thick, ->] (flycheck.east) -- (src3.west);

  \draw [thick, ->] (src1.east) to (src-db.west);
  \draw [thick, ->] (src2.east) -- node [yshift=0.2mm,rotate=-36]
  {\mbox{\begin{minipage}{0.05\textwidth}\sf $\Delta$ \\ \texttt{chars}1 
      \end{minipage}
    }} (src-db.west);

  \draw [thick, ->] (src3.east) -- node [yshift=0mm]
  {\begin{minipage}{0.05\textwidth}{\sf $\Delta$ \texttt{chars}2}
    \end{minipage}
  } (src-db.west);

  \draw [thick, ->] (src-db) -- (code);

  
  \node (code-db) [ciaopp-db,right of=code,xshift=6.5em,yshift=3mm] {\textbf{CHC DB}};

\node (lib-db) [ciaopp-db,below of=code-db,yshift=3mm] {\textbf{Libraries DB}};
\node (normalizer) [below of=lib-db,ciaopp-db] {\textbf{Assertion DB}};
\node (statana) [above of=code-db,yshift=1em,tool] {Static Analyzer};

\node (anainfo) [ciaopp-db,right of=statana,xshift=5em] {\textbf{Analysis DB}};
\node (ctchecker) [tool,right of=normalizer,xshift=5em] {Static Checker};
\path [draw, thick, ->] (anainfo) -- node [] {} (ctchecker) ;
\node (true)      [color=green!50!black,above of=ctchecker,midresult] {:- true};

\path[color=blue] (ctchecker)+(7.5em,2em) node (check) [midresult] {:- check};
\path[color=red] (ctchecker)+(7.5em,-0em) node (false) [midresult] {:- false};
\path[color=green!50!black] (ctchecker)+(7.5em,-2em) node (checked) [midresult] {:- checked};
\node (spec) [tool,above of=check,yshift=3mm] {Dynamic Annotator};

\node (srct) [file,right of=spec,xshift=3em,yshift=1em] {RT safe src};
\node (rterror) [errresult,above of=srct,yshift=1em] {Possible\\run-time error};
\path (ctchecker)+(14.5em,0em) node (cterror) [errresult] {Error};
\path (ctchecker)+(14.5em,2em) node (verifwarn) [warnresult] {Warning};
\path (ctchecker)+(14.5em,-2em) node (verified) [okresult] {Verified};
\node (verified-side) [right of=verified,xshift=2mm] {};

\path [draw, thick, ->] (code-db.north) -- node [xshift=1em] {{\sf $\Delta$}} (statana.south) ;
\path [draw, thick, ->] (code-db.east)++(0,0.5em) -- node [] {} (spec) ;
\path [draw, thick, ->] (code.east) -- node [xshift=0.5em,yshift=0.2em,rotate=15]
{\begin{minipage}{0.05\textwidth}{\centering \sf $\Delta$ CHC}
  \end{minipage}
} (code-db.west) ;
\path [draw, thick, ->] (code.east) -- node [] {} (normalizer.west) ;
\draw [thick, ->] (lib-db.east) to [bend right] (statana.south east) ;
\path [draw, thick, ->] (normalizer) -- node [] {} (ctchecker) ;
\path [draw, thick, <->] (statana) -- node [] {} (anainfo.west) ;
\path [draw, thick, ->] (ctchecker.east) -- node [] {} (false.west) ;
\path [draw, thick, ->] (ctchecker.east) -- node [] {} (check.west) ;
\path [draw, thick, ->] (ctchecker.east) -- node [] {} (checked.west) ;
\path [draw, thick, ->] (false.east) -- node [] {} (cterror.west) ;
\path [draw, thick, ->] (check.east) -- node [] {} (verifwarn.west) ;
\path [draw, thick, ->] (checked.east) -- node [] {} (verified.west) ;

\draw [draw, thick, ->] (check.north) to (spec.south) ;
\path [draw, thick, ->] (spec.east) -- node [] {} (srct.west) ;
\path [draw, thick, ->] (srct) -- node [] {} (rterror.south) ;



\node (a) [below of=verified,yshift=3mm] {};
\path [draw, ultra thick, red!50] (verified.south) -- (a.north);
\path [draw, ultra thick, red!50, ->] (a.north) -| (ciao-flycheck.south);

\node (b) [right of=rterror,xshift=3mm] {};
\path [draw, ultra thick, red!50] (rterror.east) -- (b);
\path [draw, ultra thick, red!50] (b.west) |- (a.north);

\path [draw, ultra thick, red!50] (src-db.south)+(0mm,-1mm) |- (a.north);

\path [color=black] (statana.north)+(0,1.5em) node (preprocessor) {\sffamily \small \textbf{CiaoPP}};
\path [color=black] (code.north)+(0,1.5em) node (program) {\sffamily \small Front-end};
\path [color=black] (flycheck.north)+(0,1.5em) node (emacs) {\sffamily \small IDE (Emacs)};
\path [color=black] (verifwarn.north)+(0,0.9em) node (report) {\sffamily \small Report};

\begin{pgfonlayer}{background}
  \path (statana.north west)+(-3.3,0.4) node (g) {};
  \path (checked.south east)+(0.5,-0.2) node (h) {};
  
  \path[fill=yellow!20,rounded corners, draw=black!50, densely dashed] (g) rectangle (h);
\end{pgfonlayer}

\begin{pgfonlayer}{background}
  \path (code.north west)+(-0.1,0.8) node (g) {};
  \path (src-db.south east)+(0.1,-0.1) node (h) {};
  
  \path[source] (g) rectangle (h);
\end{pgfonlayer}

\begin{pgfonlayer}{background}
  \path (flycheck.north west)+(-0.1,0.8) node (g) {};
  \path (src3.south east)+(0.1,-0.5) node (h) {};
  
  \path[emacs] (g) rectangle (h);
\end{pgfonlayer}

\begin{pgfonlayer}{background}
  \path (verifwarn.north west)+(-0.1,0.5) node (g) {};
  \path (verified.south east)+(0.1,-0.1) node (h) {};
  
  \path[rounded corners, draw=black!50, densely dashed] (g) rectangle (h);
\end{pgfonlayer}

\end{tikzpicture}

  }
  \figend  
  \figend  
  \caption{Integration of the \ciaopp framework in the Emacs-based
    IDE.}\label{fig:ciaopp+ide}
  \figend  
\end{figure}

Fig.~\ref{fig:ciaopp+ide} shows the overall architecture of \verifly,
the integration of the \ciaopp framework with the new IDE components,
represented by the new box to the left, and the communication to and
from \ciaopp.
As mentioned before, the tool interface is implemented within Emacs
and the on-the-fly support is provided by the Emacs ``\flycheck''
package.\footnote{\url{https://github.com/flycheck/flycheck}}
\textsf{Flycheck} is an extension developed for GNU Emacs originally
designed for on-the-fly \emph{syntax} checking, but we use it here in
a semantic context.
However, as also mentioned before, a similar integration is possible
with any reasonably extensible IDE.

The overall architecture consists of a \flycheck adaptor (implementing
different \ciao-based checkers, from syntactic to full analysis), a
\ciaopp process that runs in the background in daemon mode, and a
lightweight client to \ciaopp.
When a file is opened, modified, or saved, as well as after some small
period of inactivity, an editor checking event is triggered. Edit
events notify \ciaopp (via the lightweight client) about program
changes, which can be both in code and assertions.
The \ciaopp daemon receives these changes, and, behind the scenes,
transforms them
into changes at the \hcir level (also checking for syntactic errors),
and then incrementally (re-)analyzes the code and (re-)checks any 
reachable assertions.
The latter can be in libraries, other modules, language built-in
specifications, or of course (but not necessarily) in user code.
The results (errors, verifications, and warnings), from static (and
possibly also dynamic checks) are returned to the IDE and presented as
colorings and tooltips directly on the program text.
This overall behavior is what we have called in our tool the
``\verifly'' mode of operation.

\parbeg
\paragraph{\textbf{Details on the architecture.}} Currently, \flycheck requires
saving the contents of the source being edited (Emacs \emph{buffer})
into a temporary file and then invoking an external command.
In our case the external command is a lightweight client that
communicates with a running \ciaopp process, executing as a an
\emph{active module}, a daemon process that executes in the background
and reacts to simple JSON-encoded queries via a
socket.\footnote{JSON-encoded interaction capabilities have been added
  recently to support tool interoperability and browser-based
  interactions, and to simplify future extensions.}
This \ciaopp process is started once and kept alive for future
analyses, ensuring that no time is unnecessarily wasted in startup and
cleanups, as well as allowing caching some common analysis data, etc.\
for libraries. The approach similar to other like LSP (Language
Server Protocol).
Finally, \ciao implements a ``shadow module'' mechanism that allows the
compiler to read alternative versions for some given modules. We use
this mechanism to make \ciaopp (and other \ciao-based checkers) read the
contents of temporary Emacs buffers during edition (saved as temporary
files with \emph{shadowed} names). This is specially useful to work in
(inter-)modular analysis where the analysis root is not necessarily
the active buffer.

\begin{figure}[t]
  \centering
  \figbeg
  \fbox{\includegraphics[width=0.60\textwidth,clip,trim=0 40 0 5]{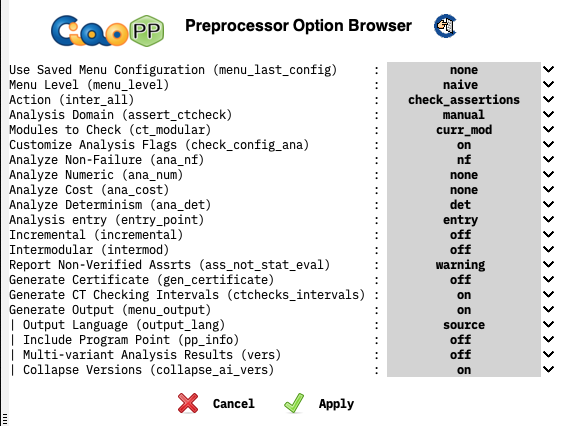}}
  \figend
  \caption{The \ciaopp option browser.}\label{fig:browser}
  \figend
\end{figure}

\parbeg
\paragraph{\textbf{Customizing the analysis.}} In general, \ciaopp can be run
fully automatically and does not \emph{require} the user to change the
configuration or provide invariants or
assertions, and, for example, selects automatically abstract domains,
as mentioned before. However, the system does have a configuration
interface that allows manual selection of abstract domains, and of
many parameters such as whether passes are incremental or not, the
level of context sensitivity, error and warning levels, the type of
output, etc.
Fig.~\ref{fig:browser} shows the option browsing
interface of the tool, as well as some options (abstract domain
selections) in the menus for the cost analysis, shape and type
analysis, pointer (logic variable) aliasing analysis, and numeric
analysis.

\begin{figure}[t]
  \centering
  \figbeg
  \fbox{\includegraphics[width=0.85\textwidth,clip,trim=0 78 4 3]{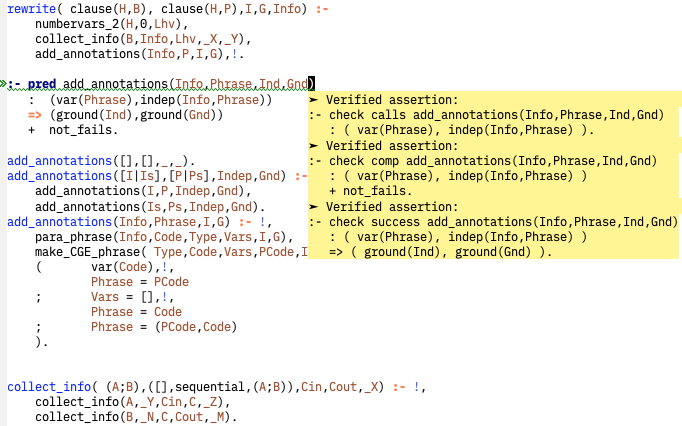}}
  \figend
  \caption{An assertion within a parallelizer (\texttt{ann}).}\label{fig:ann}
  \figend
\end{figure}

\begin{figure}
  \centering
  \figbeg
  \fbox{\includegraphics[width=0.94\textwidth,clip,trim=0 25 0 0]{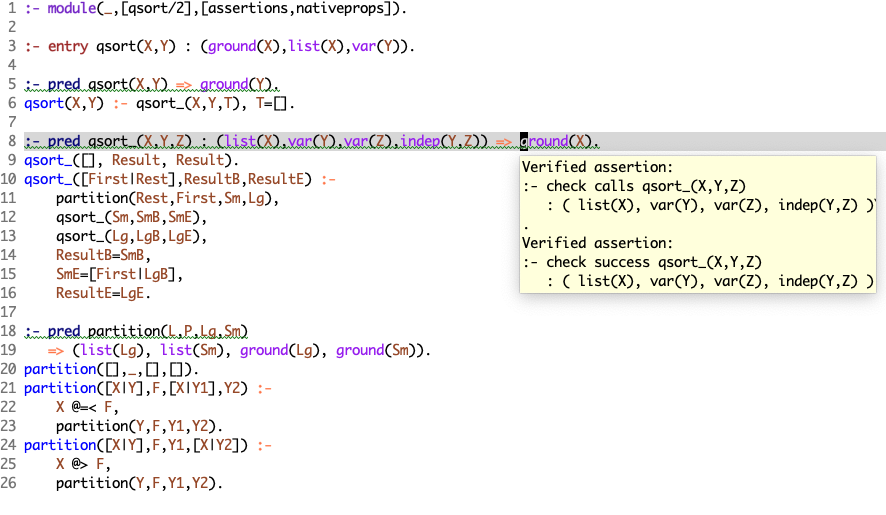}}
  \figend
  \caption{Sorting with incomplete data structures.}\label{fig:qsortdl-verified} 
  \figend
\end{figure}

\parbeg
\paragraph{\textbf{\verifly in action:}}
We now show some simple examples of the system in action.
Fig.~\ref{fig:ann} shows an
assertion being verified within a medium-sized program implementing an
automatic program parallelizer. The \texttt{add\_annotations} loop
traverses recursively a list of blocks and transforms sequential
sections into parallel expressions.  Upon opening the file the
assertion is underlined in green, meaning that it has been verified
({\tt \achecked} status).  This ensures that upon entering the
procedure there is no variable (pointer) sharing between \texttt{Info}
(the input) and \texttt{Phrase}, i.e., \texttt{indep(Info,Phrase)};
that \texttt{Phrase} will arrive always as a free variable; and that
on output from the procedure, \texttt{Ind} and \texttt{Gnd} will be
ground terms (i.e., will contain no null pointers).
Furthermore, this procedure is guaranteed not to fail. The
corresponding information is also highlighted in the tool-tip (in
yellow).

In Fig.~\ref{fig:qsortdl-verified} 
we show
an implementation of quick-sort using open-ended (``difference'') lists to construct the output
lists (i.e., using pointers to append in constant time). 

\begin{figure}[t]
  \centering
  \figbeg
  \fbox{\includegraphics[width=0.75\textwidth,clip,trim=1 25 38 2]{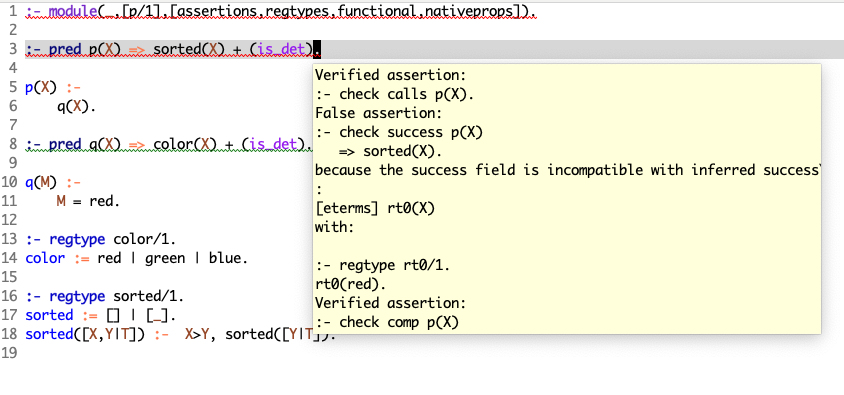}}
  \figend
  \caption{A property incompatibility bug detected statically.}\label{fig:simplebug}
  \figend
\end{figure}

\begin{figure}[t]
  \begin{tabular}{cc}
    \begin{subfigure}[a]{0.5\linewidth}
      \includegraphics[width=1\textwidth,clip,trim=1 0 20 2]{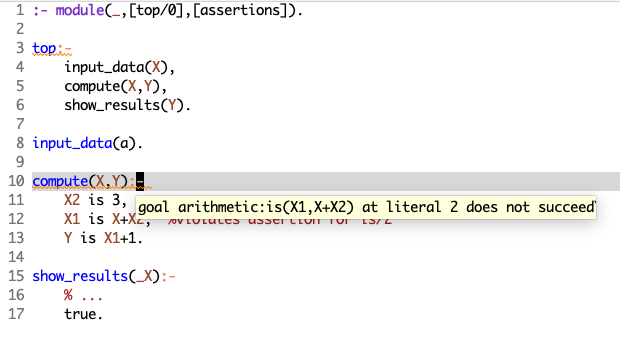}
      \figend
      \figend
      \figend
      \caption{Static detection of illegal call to lib 
        predicate.
        \figend
      }\label{fig:builtin}
    \end{subfigure}
    &
      \begin{subfigure}[a]{0.5\linewidth}
        \includegraphics[width=1\textwidth,clip,trim=1 0 25 0]{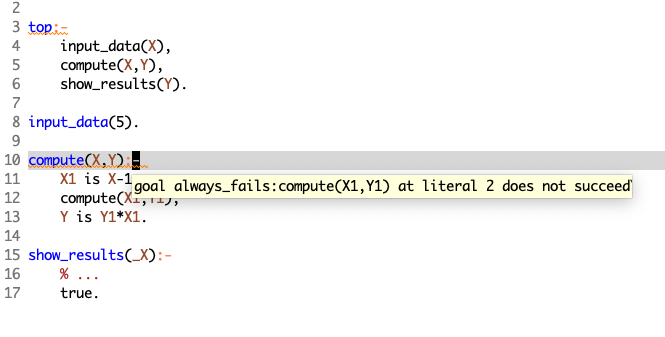}
        \figend
        \figend
        \caption{Static detection of simple non-termination.}\label{fig:does-not-succeed}
      \end{subfigure}
  \end{tabular}
      \figend
  \caption{Detection of errors at program points.}
  \figend
\end{figure}

\begin{figure}[t]
  \centering
  \fbox{\includegraphics[width=1.0\textwidth,clip,trim=1 0 0 0]{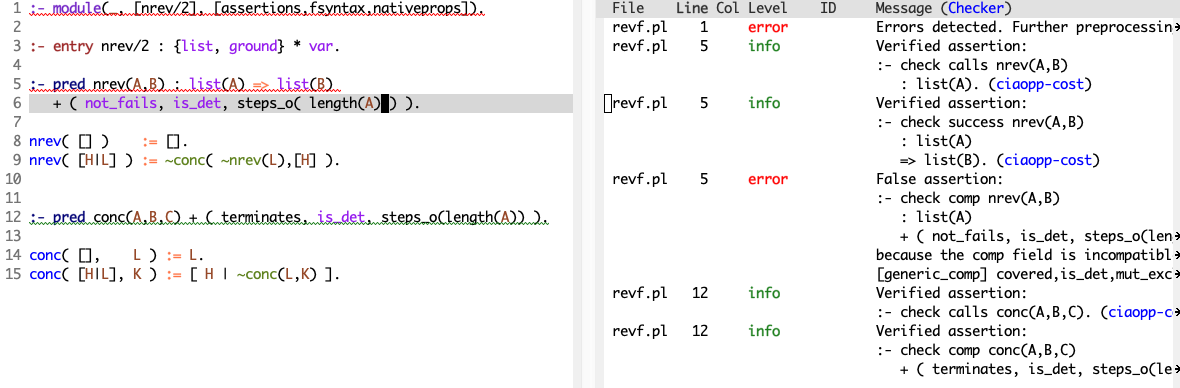}}
  \figend
  \figend
  \caption{Static verification of determinacy, termination, and cost (errors detected).}\label{fig:cost-errors}
  \figend
\end{figure}

\begin{figure}[b]
  \centering
  \fbox{\includegraphics[width=1.0\textwidth,clip,trim=1 0 25 0]{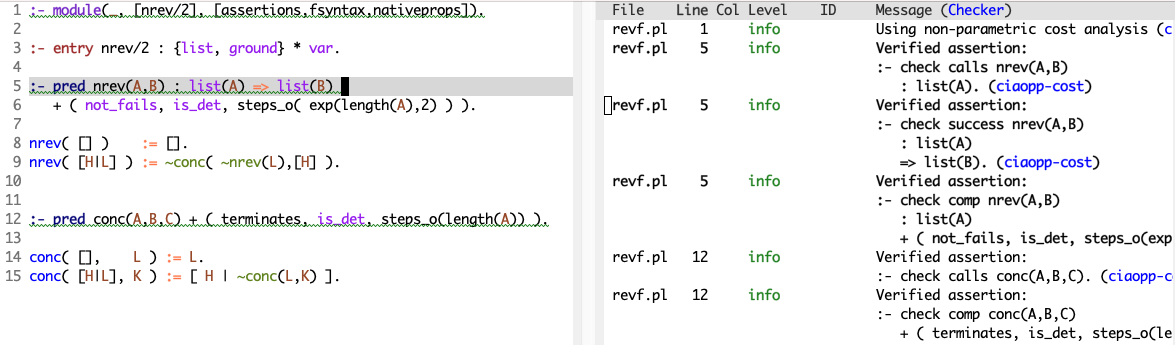}}
   \figend
   \figend
  \caption{Static verification of determinacy, termination, and cost (verified).}\label{fig:cost-verified}
   \figend
   \figend
\end{figure}

Examples~\ref{fig:simplebug},~\ref{fig:builtin},
and~\ref{fig:does-not-succeed} show static detection of, respectively,
a property incompatibility bug (note that the concrete property
\texttt{sorted/1} is approximated by the regtypes domain as
\texttt{list/1} and an incompatibility is found with \texttt{rt0},
i.e., \texttt{red}), an illegal call to a library predicate, and
a simple non-termination.

Fig.~\ref{fig:cost-errors} shows again the naive reverse
example of Fig.~\ref{fig:nrevf}, modified to 
illustrate the detection of an error
regarding unintended behavior w.r.t.\ cost.  The assertion in lines 5
and 6 of Fig.~\ref{fig:cost-errors} (left) states that predicate
\texttt{nrev} should be linear in the length of the (input) argument
\texttt{A}. This is expressed by the property
\texttt{steps\_o(length(A))}, meaning that the cost, in terms of
resolution steps, of any call to \texttt{nrev(A, B)} with \texttt{A}
bound to a list and \texttt{B} a free variable, is in
$O($\texttt{length(A)}$)$.  However, this worst case asymptotic
complexity stated in the user-provided assertion is incompatible with
a safe, quadratic lower bound on the cost of such calls ($ \frac{1}{2}
\ length{(A)}^2 + \frac{3}{2} \ length(A) + 1$) inferred by the static
analyzer. In contrast, assertion in lines 5 and 6 of
Fig.~\ref{fig:cost-verified} (left) states that predicate
\texttt{nrev} should have a quadratic worst case asymptotic complexity
in the length of \texttt{A}, which is proved to hold by means of the
upper-bound cost function inferred by analysis (which coincides with
the lower bound above). Fig.~\ref{fig:cost-verified} also shows the
verification of determinacy, non-failure, and termination properties.

\secbeg
\section{Some Performance Results}\label{sec:results}
\secend

\newcommand{\sondom}{\texttt{pairSh}}
\newcommand{\defdom}{\texttt{def}}
\newcommand{\shfrcldom}{\texttt{ShGrC}}

\begin{table}[t]
  \caption{Average response time (seconds) for the experiments with
    any program edit.}\label{tab:exp-any}
    \figend
  \begin{minipage}{\textwidth}
    \begin{tabular}{rrrcrrcrrc}
      \toprule
      & \multicolumn{3}{c}{\texttt{aggreg}} & \multicolumn{3}{c}{\texttt{readin}} & \multicolumn{3}{c}{\texttt{talkr}} \\
      \cmidrule(lr){2-4} \cmidrule(lr){5-7} \cmidrule(lr){8-10}
      domain & \textit{noinc} & \textit{inc} & \textit{speedup} & \textit{noinc} & \textit{inc} & \textit{speedup}  & \textit{noinc} & \textit{inc} & \textit{speedup} \\
      \midrule
      \sondom & 2.8 & 1.6 & $\times 1.8$ & 2.9 & 1.5 & $\times 1.9$ & 2.8 & 1.6 & $\times 1.8$ \\
      \defdom &  3.0 & 1.6 & $\times 1.9$ & 2.7 & 1.5 & $\times 1.8$ & 2.9 & 1.7 & $\times 1.7$ \\ 
      \shfrcldom & 18.1 & 5.1 & $\mathbf{\times 3.5}$ & 18.3 & 5.1 & $\mathbf{\times 3.6}$ & 18.1 & 4.5 & $\mathbf{\times 4.0}$ \\
      \bottomrule
    \end{tabular}
  \end{minipage}
  \caption{Average response time (seconds) for the experiments only
    changing assertions.}\label{tab:exp-assertions}
  \begin{minipage}{\textwidth}
    \begin{tabular}{rccccccccc}
      \toprule
      & \multicolumn{3}{c}{\texttt{aggreg}} & \multicolumn{3}{c}{\texttt{readin}} & \multicolumn{3}{c}{\texttt{talkr}} \\
      \cmidrule(lr){2-4} \cmidrule(lr){5-7} \cmidrule(lr){8-10}
      domain & \textit{noinc} & \textit{inc} & \textit{speedup} & \textit{noinc} & \textit{inc} & \textit{speedup}  & \textit{noinc} & \textit{inc} & \textit{speedup} \\
      \midrule
      \sondom & 2.8 & 1.7 & $\times 1.6$ &  2.7 & 1.6 & $\times 1.7$ & 2.9 & 1.7 & $\times 1.7$ \\
      \defdom   & 3.1 & 1.5 & $\times 2.0$ &  2.9 & 1.4 & $\times 2.0$ & 3.0 & 1.6 & $\times 1.9$ \\
      \shfrcldom  & 18.2 & 2.0 & $\mathbf{\times 9.1}$ &  18.1 & 1.9 & $\mathbf{\times 9.6}$ & 18.2 & 1.9 & $\mathbf{\times 9.6}$ \\
      \bottomrule
    \end{tabular}
  \end{minipage}
  \figend
\end{table}

We provide some performance results
from our tool using 
the well-known \texttt{chat-80}
program
(\url{https://github.com/ciao-lang/chat80})
which contains $5.2$k
lines of Prolog code across 27 files, and uses a number of system
libraries containing different Prolog built-ins and library predicates. 
The experiments consisted in
opening a specific module in the IDE, and activating the checking of assertions
with global analysis, i.e., analyzing the whole application as well as
the libraries, and then
performing a series of small edits, observing the total response time,
i.e., the time from edit to graphical update of assertion coloring in the IDE. 
Concretely, we performed two kinds of edits, predicate and assertions
(\textbf{E1}), and only assertions (\textbf{E2}). The edits were
performed on 
three selected files: \texttt{aggreg}, \texttt{readin}, and \texttt{talkr}.
To study whether incrementality improves response times significantly, we
included experiments enabling and disabling it. The experiments were performed
using \ciao 1.20 on a MacBook Air with the Apple M1 chip and 16 GB of RAM.\@
We evaluated the tool with three well-known abstract domains: a classic pair
sharing~\cite{MarriottSondergaard93} (\sondom), a dependency tracking
via propositional clauses domain~\cite{free-def-comb-short} (\defdom), and
sharing+groundness with clique-based widening~\cite{shcliques-padl06-shorter} (\shfrcldom).
The latter is the most precise, and, hence, the most expensive.
We used sharing/groundness domains because they are known to be costly
and at the same time, beyond mode inference, 
necessary to ensure correctness of most other analyses in (C)LP
systems that support logical variables, and furthermore in any
language that has pointers (aliasing).

Tables~\ref{tab:exp-any} and~\ref{tab:exp-assertions} show the response times of
analyzing and checking assertions in experiments \textbf{E1} and \textbf{E2} respectively. For
each of the files that were modified the table shows three columns:
\textit{noinc} is the response time in the non-incremental analysis setting,
\textit{inc} is the response time in the incremental setting, and
\textit{speedup} is the speedup of \textit{inc} vs.\ \textit{noinc}. Each of
the rows in the table correspond to each of the abstract domains. The reported
time is the average of \emph{total} roundtrip assertion checking time,
\emph{measured from the IDE}, that is, what the programmer actually perceives.

In all of the experiments incrementality provides significant
speedups.
In the first experiment, \textbf{E1} (Table~\ref{tab:exp-any}), for the \sondom\ and
\defdom\ domains, it allows keeping the response time, crucial for the
interactive use case of the analyzer that we propose, under 2~s.
For \shfrcldom, a significant speedup is also achieved (more than $\times
3.5$). The response time is borderline at $5~s$. 
The reason for this is, as mentioned, that the domain is more precise,
an thus more computationally expensive, which in fact allows
(dis)proving more assertions.
For the \textbf{E2} experiment (Table~\ref{tab:exp-assertions}), the tool detects that
no changes are required in the analysis results, and the assertions can be
rechecked w.r.t.\ the available (previous) analysis. The performance analyzing
with \shfrcldom\ is improved significantly (more than $\times 9$) but, more
importantly, the response times are around $2~s$.
All in all, the incremental features allow using many domains and, at
least in some cases, even the most expensive domains.

Note that the experiments reveal also that an interesting
configuration of this tool is to run different analyses in a
portfolio, where which analyses to run is decided depending on the
kind of change occurred.
If only assertions have changed, it is enough to recheck only with
\shfrcldom.  However if both the code and the assertions changed,
analysis for all domains can be run in parallel giving fast, less
precise feedback to the programmer as the results are available, and
then refine the results when the more precise results are ready.

Aside from the data in the tables, we observed a constant overhead of
$0.4$~s for loading the code---parsing and prior transformations---in the
tool.
This is currently still not fully incremental and has not been
optimized yet to load only the parts that change.
Verification times are negligible w.r.t.\ analysis times and are
approx.\ $0.1$~s; this is also non incremental, since we found that it
is not currently a bottleneck, although we plan to make it incremental
in the future to push the limits of optimizations forward.

\secbeg
\section{Related Work}\label{sec:related}
\secend

The topic of assertion checking in logic programming, and in Prolog in
particular, has received considerable attention. A family of
approaches involves defining static type systems for logic
programs~\cite{mycroft84:type_system_prolog,LakshmanReddy91,Pfenning92TILP} 
and several strongly-typed logic programming systems have been
proposed, notable examples being Mercury~\cite{mercury-jlp-short} and
G{\"o}del~\cite{goedel-short}. Approaches for combining strongly typed
and untyped Prolog modules were proposed
in~\cite{schrijvers08:typed_prolog-short}.
Most of these approaches impose a number of restrictions
that make them less appropriate for dynamic languages like Prolog.
The \ciao model introduced an alternative for writing safe programs
without relying on full static typing, but based instead the notions
of safe approximations and abstract interpretation, providing a more
general and flexible approach than in previous work, since assertions
are optional and can contain any abstract property.  This approach is
specially useful for dynamic languages
---see,
e.g.,~\citeauthor{hermenegildo11:ciao-stop-short}~(\citeyear{hermenegildo11:ciao-stop-short})
for a discussion of this topic.
Some aspects of the \ciao model have been adopted or applied in other
recent Prolog-based approaches, such as,
e.g.,~\citeauthor{schrijvers08:typed_prolog-short}~(\citeyear{schrijvers08:typed_prolog-short}),~\citeauthor{koerner-types-prolog-wflp20-shorter}~(\citeyear{koerner-types-prolog-wflp20-shorter})
or the library for run-time checking of assertions in SWI-Prolog.
It can be considered an antecedent of the now popular
\emph{gradual-} and \emph{hybrid-typing}
approaches~\cite{DBLP:conf/popl/Flanagan06-hybrid-type-checking-shortest,Siek06gradualtyping,DBLP:conf/popl/RastogiSFBV15-short}.
Moreover, recent
work~\cite{optchk-journal-scp-short,termhide-padl2018-short}
illustrates further that these techniques and the exploitation of
local data invariants allows maintaining strong safety guarantees
with zero or very low (and potentially
guaranteed~\cite{rtchecks-cost-2018-ppdp-short}) run-time check
overheads, even in a reusable-library context, without requiring
switching to strong typing. 
This approach is also more natural in Prolog-style languages that
distinguish between erroneous and failed (backtracking) executions.
All these techniques taken together strongly suggest that using the
Ciao approach comparable safety guarantees to those of strong typing
can be achieved without its requirements.
There has been work on incrementality within theorem
proving-based verification approaches, such as,
e.g.,~\citeauthor{leino-caching-verification-2015-short}~\citeyear{leino-caching-verification-2015-short}
and~\citeauthor{tschannen-oop-verification-11-short}~\citeyear{leino-caching-verification-2015-short}.
Additional references can be found in the \ciaopp overview papers and
the other citations provided.

\secbeg
\section{Conclusions}\label{sec:conclude}
\secend

We have shown how a combination of the \ciaopp static analysis and
verification framework within an integrated development environment
(IDE), can take advantage of 
incrementality 
to achieve a high level of reactivity
when reflecting analysis and verification results back as colorings
and tooltips directly on the program text. We have termed this mode of
operation ``\verifly mode.''
Our initial experience with this integrated tool shows quite promising
results, with low latency times that provide early, continuous, and
precise ``on-the-fly'' 
semantic feedback to programmers during the development process.
This allows detecting many types of errors
including swapped variables, property incompatibilities, illegal calls
to library predicates, violated numeric constraints, unintended
behavior w.r.t.\ termination, resource usage, determinism, covering
and failure,
etc.
While 
presented using the Emacs and the \flycheck package, we argue that
our 
techniques and results should be applicable to any \verifly-style
integration into a modern extensible IDE.\@
We plan to continue our work 
to achieve further reactivity and scalability improvements, enhanced
presentations of verification results, and improved diagnosis,
contributing to further improve the programming environments
available to the (C)LP programmer.

\secbeg

\bibliographystyle{acmtrans}

\small 
\begin{thebibliography}{}

\bibitem[\protect\citeauthoryear{Bueno, Cabeza, Hermenegildo, and Puebla}{Bueno
  et~al\mbox{.}}{1996}]{full-prolog-esop96-short}
{\sc Bueno, F.}, {\sc Cabeza, D.}, {\sc Hermenegildo, M.~V.}, {\sc and} {\sc
  Puebla, G.} 1996.
\newblock {G}lobal {A}nalysis of {S}tandard {P}rolog {P}rograms.
\newblock In {\em ESOP}.

\bibitem[\protect\citeauthoryear{Bueno, Carro, Hermenegildo, Lopez-Garcia, and
  (Eds.)}{Bueno et~al\mbox{.}}{2021}]{ciao-reference-manual-1.20-short}
{\sc Bueno, F.}, {\sc Carro, M.}, {\sc Hermenegildo, M.~V.}, {\sc Lopez-Garcia,
  P.}, {\sc and} {\sc (Eds.), J.~M.} 2021.
\newblock {T}he {C}iao {S}ystem. {R}ef. {M}anual (v1.20).
\newblock Tech. rep. April.
\newblock Available at \texttt{http://ciao-lang.org}.

\bibitem[\protect\citeauthoryear{Bueno, Deransart, Drabent, Ferrand,
  Hermenegildo, Maluszynski, and Puebla}{Bueno
  et~al\mbox{.}}{1997}]{aadebug97-informal-short}
{\sc Bueno, F.}, {\sc Deransart, P.}, {\sc Drabent, W.}, {\sc Ferrand, G.},
  {\sc Hermenegildo, M.~V.}, {\sc Maluszynski, J.}, {\sc and} {\sc Puebla, G.}
  1997.
\newblock {O}n the {R}ole of {S}emantic {A}pproximations in {V}alidation and
  {D}iagnosis of {C}onstraint {L}ogic {P}rograms.
\newblock In {\em Proc.\ of the 3rd Int'l. WS on Automated Debugging--AADEBUG}.
  U.\ Link\"oping Press, 155--170.

\bibitem[\protect\citeauthoryear{Cousot and Cousot}{Cousot and
  Cousot}{1977}]{Cousot77}
{\sc Cousot, P.} {\sc and} {\sc Cousot, R.} 1977.
\newblock {A}bstract {I}nterpretation: {A} {U}nified {L}attice {M}odel for
  {S}tatic {A}nalysis of {P}rograms by {C}onstruction or {A}pproximation of
  {F}ixpoints.
\newblock In {\em {ACM} {S}ymposium on {P}rinciples of {P}rogramming
  {L}anguages (POPL'77)}. ACM Press, 238--252.

\bibitem[\protect\citeauthoryear{{De Angelis}, Fioravanti, Gallagher,
  Hermenegildo, Pettorossi, and Proietti}{{De Angelis}
  et~al\mbox{.}}{2021}]{anal-peval-horn-verif-2021-tplp}
{\sc {De Angelis}, E.}, {\sc Fioravanti, F.}, {\sc Gallagher, J.~P.}, {\sc
  Hermenegildo, M.~V.}, {\sc Pettorossi, A.}, {\sc and} {\sc Proietti, M.}
  2021.
\newblock {A}nalysis and {T}ransformation of {C}onstrained {H}orn {C}lauses for
  {P}rogram {V}erification.
\newblock {\em Theory and Practice of Logic Programming\/}.

\bibitem[\protect\citeauthoryear{{De Angelis}, Fioravanti, Pettorossi, and
  Proietti}{{De Angelis} et~al\mbox{.}}{2017}]{AngelisFPP17}
{\sc {De Angelis}, E.}, {\sc Fioravanti, F.}, {\sc Pettorossi, A.}, {\sc and}
  {\sc Proietti, M.} 2017.
\newblock Semantics-based generation of verification conditions via program
  specialization.
\newblock {\em Sci. Comput. Program.\/}~{\em 147}, 78--108.

\bibitem[\protect\citeauthoryear{Dumortier, Janssens, Simoens, and {Garc\'{\i}a
  de la Banda}}{Dumortier et~al\mbox{.}}{1993}]{free-def-comb-short}
{\sc Dumortier, V.}, {\sc Janssens, G.}, {\sc Simoens, W.}, {\sc and} {\sc
  {Garc\'{\i}a de la Banda}, M.} 1993.
\newblock {C}ombining a {D}efiniteness and a {F}reeness {A}bstraction for {CLP}
  {L}anguages.
\newblock In {\em Workshop on LP Synthesis and Transformation}.

\bibitem[\protect\citeauthoryear{Flanagan}{Flanagan}{2006}]{DBLP:conf/popl/Flanagan06-hybrid-type-checking-shortest}
{\sc Flanagan, C.} 2006.
\newblock {H}ybrid {T}ype {C}hecking.
\newblock In {\em 33rd {ACM} {POPL}}. 245--256.

\bibitem[\protect\citeauthoryear{Gallagher, Hermenegildo, Kafle, Klemen,
  Lopez-Garcia, and Morales}{Gallagher
  et~al\mbox{.}}{2020}]{big-small-step-vpt2020-shorter}
{\sc Gallagher, J.}, {\sc Hermenegildo, M.~V.}, {\sc Kafle, B.}, {\sc Klemen,
  M.}, {\sc Lopez-Garcia, P.}, {\sc and} {\sc Morales, J.} 2020.
\newblock From big-step to small-step semantics and back with interpreter
  specialization.
\newblock In {\em VPT 2020}. EPTCS. Open Publishing Association, 50--65.

\bibitem[\protect\citeauthoryear{Garcia-Contreras, Morales, and
  Hermenegildo}{Garcia-Contreras
  et~al\mbox{.}}{2020}]{incanal-assrts-openpreds-lopstr19-post-short}
{\sc Garcia-Contreras, I.}, {\sc Morales, J.}, {\sc and} {\sc Hermenegildo,
  M.~V.} 2020.
\newblock {I}ncremental {A}nalysis of {L}ogic {P}rograms with {A}ssertions and
  {O}pen {P}redicates.
\newblock In {\em LOPSTR'19}. LNCS, vol. 12042. Springer, 36--56.

\bibitem[\protect\citeauthoryear{Garcia-Contreras, Morales, and
  Hermenegildo}{Garcia-Contreras
  et~al\mbox{.}}{2021}]{intermod-incanal-2020-tplp-short}
{\sc Garcia-Contreras, I.}, {\sc Morales, J.~F.}, {\sc and} {\sc Hermenegildo,
  M.~V.} 2021.
\newblock {I}ncremental and {M}odular {C}ontext-sensitive {A}nalysis.
\newblock {\em TPLP\/}~{\em 21,\/}~2 (January), 196--243.

\bibitem[\protect\citeauthoryear{G\'{o}mez-Zamalloa, Albert, and
  Puebla}{G\'{o}mez-Zamalloa et~al\mbox{.}}{2009}]{mod-decomp-jist09-short}
{\sc G\'{o}mez-Zamalloa, M.}, {\sc Albert, E.}, {\sc and} {\sc Puebla, G.}
  2009.
\newblock {D}ecompilation of {J}ava {B}ytecode to {P}rolog by {P}artial
  {E}valuation.
\newblock {\em JIST\/}~{\em 51}, 1409--1427.

\bibitem[\protect\citeauthoryear{Grebenshchikov, Gupta, Lopes, Popeea, and
  Rybalchenko}{Grebenshchikov
  et~al\mbox{.}}{2012}]{DBLP:conf/tacas/GrebenshchikovGLPR12-short}
{\sc Grebenshchikov, S.}, {\sc Gupta, A.}, {\sc Lopes, N.~P.}, {\sc Popeea,
  C.}, {\sc and} {\sc Rybalchenko, A.} 2012.
\newblock {HSF(C)}: {A} {S}oftware {V}erifier {B}ased on {H}orn {C}lauses.
\newblock In {\em TACAS}. 549--551.

\bibitem[\protect\citeauthoryear{Gurfinkel, Kahsai, Komuravelli, and
  Navas}{Gurfinkel et~al\mbox{.}}{2015}]{DBLP:conf/cav/GurfinkelKKN15-shorter}
{\sc Gurfinkel, A.}, {\sc Kahsai, T.}, {\sc Komuravelli, A.}, {\sc and} {\sc
  Navas, J.~A.} 2015.
\newblock {T}he {S}ea{H}orn {V}erification {F}ramework.
\newblock In {\em CAV}. 343--361.

\bibitem[\protect\citeauthoryear{Henriksen and Gallagher}{Henriksen and
  Gallagher}{2006}]{HGScam06-short}
{\sc Henriksen, K.~S.} {\sc and} {\sc Gallagher, J.~P.} 2006.
\newblock {A}bstract {I}nterpretation of {PIC} {P}rograms through {L}ogic
  {P}rogramming.
\newblock In {\em SCAM~'06}. IEEE Computer Society, 184--196.

\bibitem[\protect\citeauthoryear{Hermenegildo, Puebla, Bueno, and
  Garcia}{Hermenegildo et~al\mbox{.}}{2005}]{ciaopp-sas03-journal-scp-short}
{\sc Hermenegildo, M.}, {\sc Puebla, G.}, {\sc Bueno, F.}, {\sc and} {\sc
  Garcia, P.~L.} 2005.
\newblock {I}ntegrated {P}rogram {D}ebugging, {V}erification, and
  {O}ptimization {U}sing {A}bstract {I}nterpretation (and {T}he {C}iao {S}ystem
  {P}reprocessor).
\newblock {\em Science of Computer Programming\/}~{\em 58,\/}~1--2, 115--140.

\bibitem[\protect\citeauthoryear{Hermenegildo, Bueno, Carro, Lopez-Garcia,
  Mera, Morales, and Puebla}{Hermenegildo
  et~al\mbox{.}}{2011}]{hermenegildo11:ciao-stop-short}
{\sc Hermenegildo, M.~V.}, {\sc Bueno, F.}, {\sc Carro, M.}, {\sc Lopez-Garcia,
  P.}, {\sc Mera, E.}, {\sc Morales, J.}, {\sc and} {\sc Puebla, G.} 2011.
\newblock {T}he {C}iao {A}pproach to the {D}ynamic vs. {S}tatic {L}anguage
  {D}ilemma.
\newblock In {\em Proc.\ Int'l.\ WS on Scripts to Programs, STOP'11}. ACM.

\bibitem[\protect\citeauthoryear{Hermenegildo, Bueno, Carro, Lopez-Garcia,
  Mera, Morales, and Puebla}{Hermenegildo
  et~al\mbox{.}}{2012}]{hermenegildo11:ciao-design-tplp}
{\sc Hermenegildo, M.~V.}, {\sc Bueno, F.}, {\sc Carro, M.}, {\sc Lopez-Garcia,
  P.}, {\sc Mera, E.}, {\sc Morales, J.}, {\sc and} {\sc Puebla, G.} 2012.
\newblock {A}n {O}verview of {C}iao and its {D}esign {P}hilosophy.
\newblock {\em Theory and Practice of Logic Programming\/}~{\em 12,\/}~1--2
  (January), 219--252.

\bibitem[\protect\citeauthoryear{Hermenegildo, Puebla, and Bueno}{Hermenegildo
  et~al\mbox{.}}{1999}]{prog-glob-an-short}
{\sc Hermenegildo, M.~V.}, {\sc Puebla, G.}, {\sc and} {\sc Bueno, F.} 1999.
\newblock {U}sing {G}lobal {A}nalysis, {P}artial {S}pecifications, and an
  {E}xtensible {A}ssertion {L}anguage for {P}rogram {V}alidation and
  {D}ebugging.
\newblock In {\em {T}he {L}ogic {P}rogramming {P}aradigm: a 25--{Y}ear
  {P}erspective}. Springer-Verlag, 161--192.

\bibitem[\protect\citeauthoryear{Hermenegildo, Puebla, Marriott, and
  Stuckey}{Hermenegildo et~al\mbox{.}}{2000}]{incanal-toplas-short}
{\sc Hermenegildo, M.~V.}, {\sc Puebla, G.}, {\sc Marriott, K.}, {\sc and} {\sc
  Stuckey, P.} 2000.
\newblock {I}ncremental {A}nalysis of {C}onstraint {L}ogic {P}rograms.
\newblock {\em ACM TOPLAS\/}~{\em 22,\/}~2 (March), 187--223.

\bibitem[\protect\citeauthoryear{Hill and Lloyd}{Hill and
  Lloyd}{1994}]{goedel-short}
{\sc Hill, P.} {\sc and} {\sc Lloyd, J.} 1994.
\newblock {\em {T}he {G}oedel {P}rogramming {L}anguage}.
\newblock {MIT} Press.

\bibitem[\protect\citeauthoryear{Klemen, Stulova, Lopez-Garcia, Morales, and
  Hermenegildo}{Klemen et~al\mbox{.}}{2018}]{rtchecks-cost-2018-ppdp-short}
{\sc Klemen, M.}, {\sc Stulova, N.}, {\sc Lopez-Garcia, P.}, {\sc Morales,
  J.~F.}, {\sc and} {\sc Hermenegildo, M.~V.} 2018.
\newblock {S}tatic {P}erformance {G}uarantees for {P}rograms with {R}un-time
  {C}hecks.
\newblock In {\em PPDP}. ACM.

\bibitem[\protect\citeauthoryear{Lakshman and Reddy}{Lakshman and
  Reddy}{1991}]{LakshmanReddy91}
{\sc Lakshman, T.} {\sc and} {\sc Reddy, U.} 1991.
\newblock Typed {P}rolog: A semantic reconstruction of the {Mycroft-O'Keefe}
  type system.
\newblock In {\em International Logic Programming Symposium}. MIT Press.

\bibitem[\protect\citeauthoryear{Lopez-Garcia, Darmawan, Klemen, Liqat, Bueno,
  and Hermenegildo}{Lopez-Garcia
  et~al\mbox{.}}{2018}]{resource-verification-tplp18-shortest}
{\sc Lopez-Garcia, P.}, {\sc Darmawan, L.}, {\sc Klemen, M.}, {\sc Liqat, U.},
  {\sc Bueno, F.}, {\sc and} {\sc Hermenegildo, M.~V.} 2018.
\newblock {I}nterval-based {R}esource {U}sage {V}erification by {T}ranslation
  into {H}orn {C}lauses and an {A}pplication to {E}nergy {C}onsumption.
\newblock {\em TPLP\/}~{\em 18,\/}~2 (March), 167--223.

\bibitem[\protect\citeauthoryear{Marriott and S{\o}ndergaard}{Marriott and
  S{\o}ndergaard}{1993}]{MarriottSondergaard93}
{\sc Marriott, K.} {\sc and} {\sc S{\o}ndergaard, H.} 1993.
\newblock Precise and efficient groundness analysis for logic programs.
\newblock Technical report 93/7, Univ. of Melbourne.

\bibitem[\protect\citeauthoryear{M\'{e}ndez-Lojo, Navas, and
  Hermenegildo}{M\'{e}ndez-Lojo
  et~al\mbox{.}}{2007}]{decomp-oo-prolog-lopstr07-short}
{\sc M\'{e}ndez-Lojo, M.}, {\sc Navas, J.}, {\sc and} {\sc Hermenegildo, M.}
  2007.
\newblock {A} {F}lexible ({C}){LP}-{B}ased {A}pproach to the {A}nalysis of
  {O}bject-{O}riented {P}rograms.
\newblock In {\em LOPSTR}. LNCS, vol. 4915. Springer-Verlag, 154--168.

\bibitem[\protect\citeauthoryear{Mycroft and O'Keefe}{Mycroft and
  O'Keefe}{1984}]{mycroft84:type_system_prolog}
{\sc Mycroft, A.} {\sc and} {\sc O'Keefe, R.~A.} 1984.
\newblock A polymorphic type system for {P}rolog.
\newblock {\em Artificial Intelligence\/}~{\em 23,\/}~3, 295--307.

\bibitem[\protect\citeauthoryear{Navas, Bueno, and Hermenegildo}{Navas
  et~al\mbox{.}}{2006}]{shcliques-padl06-shorter}
{\sc Navas, J.}, {\sc Bueno, F.}, {\sc and} {\sc Hermenegildo, M.~V.} 2006.
\newblock {Efficient Top-Down Set-Sharing Analysis Using Cliques}.
\newblock In {\em 8th PADL}. Number 2819 in LNCS. Springer, 183--198.

\bibitem[\protect\citeauthoryear{Peralta, Gallagher, and Sa{\v{g}}lam}{Peralta
  et~al\mbox{.}}{1998}]{Peralta-Gallagher-Saglam-SAS98}
{\sc Peralta, J.}, {\sc Gallagher, J.}, {\sc and} {\sc Sa{\v{g}}lam, H.} 1998.
\newblock Analysis of imperative programs through analysis of constraint logic
  programs.
\newblock In {\em Static Analysis. 5th International Symposium, SAS'98, Pisa},
  {G.~Levi}, Ed. LNCS, vol. 1503. 246--261.

\bibitem[\protect\citeauthoryear{Perez-Carrasco, Klemen, Lopez-Garcia, Morales,
  and Hermenegildo}{Perez-Carrasco
  et~al\mbox{.}}{2020}]{resources-blockchain-sas20-short}
{\sc Perez-Carrasco, V.}, {\sc Klemen, M.}, {\sc Lopez-Garcia, P.}, {\sc
  Morales, J.}, {\sc and} {\sc Hermenegildo, M.~V.} 2020.
\newblock {C}ost {A}nalysis of {S}mart {C}ontracts via {P}arametric {R}esource
  {A}nalysis.
\newblock In {\em Static Aanalysis Simposium (SAS'20)}. LNCS, vol. 12389.
  Springer, 7--31.

\bibitem[\protect\citeauthoryear{Pfenning}{Pfenning}{1992}]{Pfenning92TILP}
{\sc Pfenning, F.}, Ed. 1992.
\newblock {\em Types in Logic Programming}.
\newblock MIT Press.

\bibitem[\protect\citeauthoryear{Pietrzak, Correas, Puebla, and
  Hermenegildo}{Pietrzak et~al\mbox{.}}{2006}]{mod-ctchecks-lpar06-short}
{\sc Pietrzak, P.}, {\sc Correas, J.}, {\sc Puebla, G.}, {\sc and} {\sc
  Hermenegildo, M.~V.} 2006.
\newblock {C}ontext-{S}ensitive {M}ultivariant {A}ssertion {C}hecking in
  {M}odular {P}rograms.
\newblock In {\em LPAR'06}. Number 4246 in LNCS. Springer-Verlag, 392--406.

\bibitem[\protect\citeauthoryear{Puebla, Bueno, and Hermenegildo}{Puebla
  et~al\mbox{.}}{2000a}]{assert-lang-disciplbook-short}
{\sc Puebla, G.}, {\sc Bueno, F.}, {\sc and} {\sc Hermenegildo, M.~V.} 2000a.
\newblock {A}n {A}ssertion {L}anguage for {C}onstraint {L}ogic {P}rograms.
\newblock In {\em {A}nalysis and {V}isualization {T}ools for {C}onstraint
  {P}rogramming}. Number 1870 in LNCS. Springer-Verlag, 23--61.

\bibitem[\protect\citeauthoryear{Puebla, Bueno, and Hermenegildo}{Puebla
  et~al\mbox{.}}{2000b}]{assrt-theoret-framework-lopstr99-short}
{\sc Puebla, G.}, {\sc Bueno, F.}, {\sc and} {\sc Hermenegildo, M.~V.} 2000b.
\newblock {C}ombined {S}tatic and {D}ynamic {A}ssertion-{B}ased {D}ebugging of
  {C}onstraint {L}ogic {P}rograms.
\newblock In {\em Proc. of LOPSTR'99}. LNCS 1817. Springer-Verlag, 273--292.

\bibitem[\protect\citeauthoryear{Puebla and Hermenegildo}{Puebla and
  Hermenegildo}{1996}]{inc-fixp-sas-short}
{\sc Puebla, G.} {\sc and} {\sc Hermenegildo, M.~V.} 1996.
\newblock {O}ptimized {A}lgorithms for the {I}ncremental {A}nalysis of {L}ogic
  {P}rograms.
\newblock In {\em SAS'96}. Springer LNCS 1145, 270--284.

\bibitem[\protect\citeauthoryear{Rastogi, Swamy, Fournet, Bierman, and
  Vekris}{Rastogi et~al\mbox{.}}{2015}]{DBLP:conf/popl/RastogiSFBV15-short}
{\sc Rastogi, A.}, {\sc Swamy, N.}, {\sc Fournet, C.}, {\sc Bierman, G.}, {\sc
  and} {\sc Vekris, P.} 2015.
\newblock {Safe {\&} Efficient Gradual Typing for TypeScript}.
\newblock In {\em {42nd POPL}}. ACM, 167--180.

\bibitem[\protect\citeauthoryear{Rustan, Leino, and W{\"{u}}stholz}{Rustan
  et~al\mbox{.}}{2015}]{leino-caching-verification-2015-short}
{\sc Rustan, K.}, {\sc Leino, M.}, {\sc and} {\sc W{\"{u}}stholz, V.} 2015.
\newblock Fine-grained caching of verification results.
\newblock In {\em CAV}, {D.~Kroening} {and} {C.~S. Pasareanu}, Eds. LNCS, vol.
  9206. Springer, 380--397.

\bibitem[\protect\citeauthoryear{Schrijvers, {Santos Costa}, Wielemaker, and
  Demoen}{Schrijvers et~al\mbox{.}}{2008}]{schrijvers08:typed_prolog-short}
{\sc Schrijvers, T.}, {\sc {Santos Costa}, V.}, {\sc Wielemaker, J.}, {\sc and}
  {\sc Demoen, B.} 2008.
\newblock {T}owards {T}yped {P}rolog.
\newblock In {\em ICLP'08}. Number 5366 in LNCS. Springer, 693--697.

\bibitem[\protect\citeauthoryear{Siek and Taha}{Siek and
  Taha}{2006}]{Siek06gradualtyping}
{\sc Siek, J.~G.} {\sc and} {\sc Taha, W.} 2006.
\newblock {G}radual {T}yping for {F}unctional {L}anguages.
\newblock In {\em Scheme and Functional Programming Workshop}. 81--92.

\bibitem[\protect\citeauthoryear{Somogyi, Henderson, and Conway}{Somogyi
  et~al\mbox{.}}{1996}]{mercury-jlp-short}
{\sc Somogyi, Z.}, {\sc Henderson, F.}, {\sc and} {\sc Conway, T.} 1996.
\newblock {T}he {E}xecution {A}lgorithm of {M}ercury: an {E}fficient {P}urely
  {D}eclarative {L}ogic {P}rogramming {L}anguage.
\newblock {\em JLP\/}~{\em 29,\/}~1--3 (October), 17--64.

\bibitem[\protect\citeauthoryear{Stulova, Morales, and Hermenegildo}{Stulova
  et~al\mbox{.}}{2018a}]{termhide-padl2018-short}
{\sc Stulova, N.}, {\sc Morales, J.~F.}, {\sc and} {\sc Hermenegildo, M.~V.}
  2018a.
\newblock {E}xploiting {T}erm {H}iding to {R}educe {R}un-time {C}hecking
  {O}verhead.
\newblock In {\em Int'l.\ Symp.\ on Practical Aspects of Declarative Languages
  (PADL)}. Number 10702 in LNCS. Springer-Verlag, 99--115.

\bibitem[\protect\citeauthoryear{Stulova, Morales, and Hermenegildo}{Stulova
  et~al\mbox{.}}{2018b}]{optchk-journal-scp-short}
{\sc Stulova, N.}, {\sc Morales, J.~F.}, {\sc and} {\sc Hermenegildo, M.~V.}
  2018b.
\newblock {S}ome {T}rade-offs in {R}educing the {O}verhead of {A}ssertion
  {R}un-time {C}hecks via {S}tatic {A}nalysis.
\newblock {\em Science of Computer Programming\/}~{\em 155}, 3--26.
\newblock Highest ranked paper.

\bibitem[\protect\citeauthoryear{Tschannen, Furia, Nordio, and Meyer}{Tschannen
  et~al\mbox{.}}{2011}]{tschannen-oop-verification-11-short}
{\sc Tschannen, J.}, {\sc Furia, C.~A.}, {\sc Nordio, M.}, {\sc and} {\sc
  Meyer, B.} 2011.
\newblock Usable verification of object-oriented programs by combining static
  and dynamic techniques.
\newblock In {\em SEFM}. LNCS, vol. 7041. Springer, 382--398.

\bibitem[\protect\citeauthoryear{Wingen and K{\"{o}}rner}{Wingen and
  K{\"{o}}rner}{2020}]{koerner-types-prolog-wflp20-shorter}
{\sc Wingen, I.} {\sc and} {\sc K{\"{o}}rner, P.} 2020.
\newblock Effectiveness of annotation-based static type inference.
\newblock In {\em WS on Functional and Constraint Logic Programming}. LNCS,
  vol. 12560. Springer, 74--93.

\end{thebibliography}

\end{document}